\documentstyle[multicol,eqsecnum,aps,psfig]{revtex}
\renewcommand{\narrowtext}{\begin{multicols}{2}
\global\columnwidth20.5pc\noindent}
\renewcommand{\widetext}{\end{multicols}
\global\columnwidth42.5pc}
\begin{document}
\draft
\preprint{2 October 2000}
%\preprint{22 November 2000}
\title{Characterization of halogen-bridged binuclear metal complexes
as hybridized two-band materials}
\author{Shoji Yamamoto}
\address{Department of Physics, Okayama University,
         Tsushima, Okayama 700-8530, Japan}
%\date{Received \hspace{6cm}}
\date{Received 2 October 2000}
%\date{Received 22 November 2000}
\maketitle
\begin{abstract}
We study the electronic structure of halogen-bridged binuclear metal
(MMX) complexes with a two-band Peierls-Hubbard model.
Based on a symmetry argument, various density-wave states are derived
and characterized.
The ground-state phase diagram is drawn within the Hartree-Fock
approximation, while the thermal behavior is investigated using a
quantum Monte Carlo method.
All the calculations conclude that a typical MMX compound
Pt$_2$(CH$_3$CS$_2$)$_4$I should indeed be regarded as a
$d$-$p$-hybridized two-band material, where the oxidation of the
halogen ions must be observed even in the ground state, whereas
another MMX family (NH$_4$)$_4$[Pt$_2$(P$_2$O$_5$H$_2$)$_4$X] may be
treated as single-band materials.
\end{abstract}
\pacs{PACS numbers: 71.10.Hf, 71.45.Lr, 75.30.Fv, 75.40.Mg}
\narrowtext
\section{Introduction}

   The family of transition-metal (M) linear-chain complexes
containing bridging halogens (X) has a long history of research
\cite{Robi17,Day91,Kell57} especially in the chemical field.
The conventional class \cite{Clar95} of these materials, which we
refer to as MX chains, is represented by the Wolffram red salts
($\mbox{M}=\mbox{Pt}, \mbox{X}=\mbox{Cl}$) and has been fascinating
both chemists and physicists due to its novel properties, such as
intense and dichroic charge-transfer absorption, strong resonance
enhancement of Raman spectra, and luminescence with large Stokes
shift, which all originate in the mixed-valence ground state.
On the other hand, the synthesis of NiX compounds
\cite{Toft61,Tori41} which exhibit mono-valence magnetic ground
states aroused a renewed interest in this system.
Substituting the metals, bridging halogens, ligand molecules, and
counter anions surrounding the one-dimensional chains, the tunability
of the ground-state electronic structure was fully revealed
\cite{Okam09}.
A pressure-induced reverse Peierls instability \cite{Swan05} was also
demonstrated.
All these observations can be understood as the competition between
the Peierls and Mott insulators \cite{Nasu65}.
Intrinsic multiband effects, together with competing
electron-electron and electron-phonon interactions, raise a
possibility of further ground states appearing, such as
incommensurate (long-period) charge density waves \cite{Bati28},
spin-Peierls states \cite{Rode98}, and density waves on the halogen
sublattice \cite{Yama22}.
A leading group \cite{Gamm08} in Los Alamos National Laboratory
presented an extensive two-band-model study covering the ground-state
properties, excitation spectra, and quantum lattice fluctuations.
The research trend is now shifting toward mixed-metal complexes
\cite{Okam61,Maru99},
where the ground-state degeneracy is lifted and therefore the
relaxation process of photogenerated excitons via solitons or
polarons can efficiently be controlled.

   In an attempt to explore further into this unique system, the new
class of these materials, which we refer to as MMX chains, has been
synthesized, where binuclear metal units are bridged by halogen ions.
Thus-far synthesized MMX-chain compounds are classified into two
groups:
M$_2$(dta)$_4$I
($\mbox{M}=\mbox{Pt},\mbox{Ni}$;
 $\mbox{dta}=\mbox{dithioacetate}=\mbox{CH}_3\mbox{CS}_2^{\,-}$)
\cite{Bell44,Bell15} and
R$_4$[Pt$_2$(pop)$_4$X]$\cdot$$n$H$_2$O
($\mbox{X}=\mbox{Cl},\mbox{Br},\mbox{I}$;
 $\mbox{R}=\mbox{K},\mbox{NH}_4,\mbox{etc.}$;
 $\mbox{pop}=\mbox{diphosphonate}
 =\mbox{H}_2\mbox{P}_2\mbox{O}_5^{\,2-}$)
\cite{Che04}, which we hereafter refer to as ``dta" and ``pop"
complexes.
The binucleation of the metal sites does much more than simply
increases the internal degrees of freedom of the electronic
configuration:

  (i) The formal oxidation state of metal ions is $2.5+$, that is
      mixed valency of M$^{2+}$ and M$^{3+}$, in MMX chains, whereas
      it is $3+$, that is, mixed valency of M$^{2+}$ and M$^{4+}$, in
      MX chains.
      There exists an unpaired electron per one metal-dimer unit of
      trapped valence M$^{2+}$-M$^{3+}$, while none at all in the
      corresponding unit of trapped valence M$^{2+}$-M$^{4+}$.

 (ii) The direct M-M overlap contributes to the reduction of the
      effective on-site Coulomb repulsion and thus electrons can be
      more itinerant in MMX chains.

(iii) In the MX-chain system, metals are tightly locked together due
      to the hydrogen bonds between the amino-groups of the ligands
      and the counter anions.
      Therefore any dimerization of the metal sublattice has not
      yet been observed.
      In the MMX-chain system, adjacent metals are locked to each
      other by the surrounding ligands, but the metal-dimer moiety
      itself is rather movable especially in the dta complexes with
      neutral chain structures.
      Although there still exist weak van der Waals contacts between
      ligands against the Peierls instability in the dta complexes,
      Pt$_2$(dta)$_4$I has recently been reported to exhibit a
      dimerization of the metal sublattice \cite{Kita11}.

 (iv) Owing to the neutral chain structure, the band filling of the
      dta complexes might be varied.
\smallskip

\noindent
Thus, the new class of the halogen-bridged metal complexes is highly
potential in every aspect.
Further efforts to synthesize general polynuclear metal complexes
\cite{Saka66} stimulate our interest all the more.

\begin{figure}
\begin{flushleft}
\ \mbox{\psfig{figure=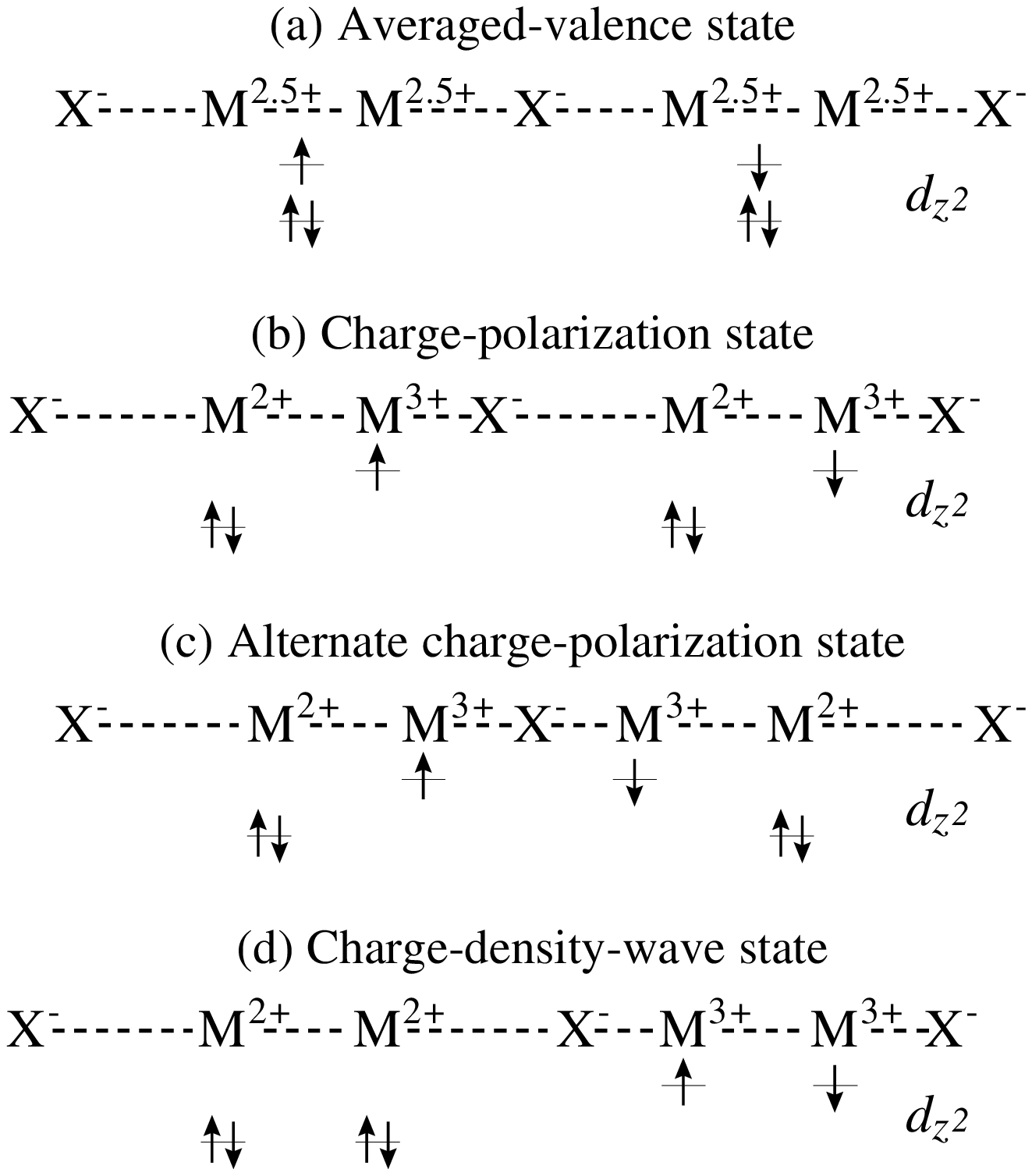,width=80mm,angle=0}}
\end{flushleft}
%\vskip 18mm
\caption{Schematic representation of electronic structures of the MMX
         chain, where the electrons in the M $d_{z^2}$ orbitals are
         regarded as efficient, while the X $p_z$ orbitals, which are
         fully occupied in the crystal rearrangement, are treated as
         irrelevant to the low-energy physics.}
\label{F:VS}
\end{figure}

   The physical properties of the MMX-chain system are not so well
established as the accumulated chemical knowledge.
There is a hot argument about the valence structures of MMX chains.
So far four types of the oxidation states have been pointed out,
which are illustrated in Fig. \ref{F:VS}.
All but (a) are accompanied by lattice distortion, where the X sites
are displaced maintaining the translational symmetry in (b), the
M$_2$ sublattice is dimerized in (c), and the X sublattice is
dimerized in (d). In this notation, (a) represents
spin-density-wave states of any kind as well as the paramagnetic
(metallic) state.
%\vskip 10mm

   The ground-state valence structure of the Cl- and Br-bridged Pt
complexes of the pop family was extensively investigated by resonance
Raman spectroscopy \cite{Kurm20}, X-ray single-crystal structure
analyses \cite{Clar09,Butl55}, solid-state $^{31}$P
nuclear-magnetic-resonance measurements \cite{Kimu40}, and polarized
optical spectra \cite{Wada95}, and was concluded to be (d).
However, due to the small Peierls gap, the ground states of these
materials can be tuned with pressure from the strongly
{\it valence-trapped} state (d) toward the {\it valence-delocalized}
state (a) (or possibly (c)) \cite{Conr23}, where the pressure mainly
contributes to the increase of the electron transfer between
neighboring Pt$_2$ moieties.
Such charge fluctuations become more remarkable when bridging
halogens are replaced by I \cite{Yama21}.
The valence transition is sensitive to the radius of the counter
cations and the number of the water molecules as well.

   For the dta complexes, on the other hand, the electronic
structures seem to be more complicated and less established.
X-ray photoelectron spectra of the Ni complexes \cite{Bell15} suggest
the valence delocalization on the regular chain structure (a), which
is convincing on the analogy of NiX chains with magnetic ground
states characterized as Mott-Hubbard insulators \cite{Okam38}.
Although early measurements \cite{Bell44,Bell15} on the Pt complexes
of the dta series implied their structural similarities with the Ni
complexes and the valence delocalization at room temperature, recent
experiments \cite{Kita68} have reported another scenario for the
valence structure of the Pt complexes:
With decreasing temperature, there occurs a metal-semiconductor
transition, that is, a transition from the averaged-valence state (a)
to the trapped-valence state (b), at $300$ K, and further transition
to the charge-ordering mode (c) follows around $80$ K.
The valence structure (c) accompanied by the metal-sublattice
dimerization is all the more interesting in relation to spin-Peierls
fluctuations, which have never yet been observed in the MX-chain
system.

   In response to such stimulative observations, we here calculate
the electronic structure of MMX compounds employing a
$\frac{5}{6}$-filled two-band model.
Under the notation of Fig. \ref{F:VS}, the filled X $p_z$ orbitals
are put out of account.
It is sometimes assumed in phenomenological arguments and
experimental investigations that unless the system is photoexcited
and/or activated by hole doping, the halogen ions contribute no
effective electron to the low-energy physics.
The goal of this paper is to give a criterion for when the
$\frac{3}{4}$-filled signle-band approximation is valid or invalid.
We calculate ground-state phase diagrams within the Hartree-Fock
approximation, where we analytically classify and characterize
possible density-wave solutions making full use of symmetry
properties.
Based on the thus-revealed ground-state properties,
we further carry out quantum Monte Carlo calculations of the thermal
properties.
Extensive analytical and numerical calculations bring us a wide view
of the $d$-$p$-hybridized two-band electron-phonon system.

\section{Hartree-Fock calculation based on a symmetry argument}

\subsection{Model Hamiltonian and Its Symmetry Properties}

   We introduce the $\frac{5}{6}$-filled one-dimensional two-band
three-orbital extended Peierls-Hubbard Hamiltonian:
\widetext
\begin{eqnarray}
   {\cal H}
   &=&\frac{K}{2}\sum_{n}
      \big(
       l_{1:n}^2+l_{2:n}^2
      \big)
    + \sum_{n,s}
      \Bigl[
       \big(\varepsilon_{\rm M}-\beta l_{1:n}\big)n_{1:n,s}
      +\big(\varepsilon_{\rm M}+\beta l_{2:n}\big)n_{2:n,s}
      +     \varepsilon_{\rm X}n_{3:n,s}
      \Bigr]
      \nonumber \\
   &-&\sum_{n,s}
      \Bigl[
       \big(t_{\rm MX}-\alpha l_{1:n}\big)a_{1:n,s}^\dagger a_{3:n,s}
      +\big(t_{\rm MX}+\alpha l_{2:n}\big)a_{2:n,s}^\dagger a_{3:n,s}
      +     t_{\rm MM}\,a_{1:n,s}^\dagger a_{2:n-1,s}
      + {\rm H.c.}
      \Bigr]
      \nonumber \\
   &+&\sum_{n}
      \big(
       U_{\rm M}\,n_{1:n,+}n_{1:n,-}
      +U_{\rm M}\,n_{2:n,+}n_{2:n,-}
      +U_{\rm X}\,n_{3:n,+}n_{3:n,-}
      \big)
      \nonumber \\
   &+&\sum_{n,s,s'}
      \big(
       V_{\rm MX}\,n_{1:n,s}n_{3:n  ,s'}
      +V_{\rm MX}\,n_{2:n,s}n_{3:n  ,s'}
      +V_{\rm MM}\,n_{1:n,s}n_{2:n-1,s'}
      \big)\,,
   \label{E:H}
\end{eqnarray}
\narrowtext
where $n_{i:n,s}=a_{i:n,s}^\dagger a_{i:n,s}$ with
$a_{i:n,s}^\dagger$ being the creation operator of an electron with
spin $s=\pm$ (up and down) for the M $d_{z^2}$ ($i=1,2$) or X $p_z$
($i=3$) orbital in the $n$th MXM unit, and
$l_{i:n}=u_{3:n}-u_{i:n}$ with $u_{i:n}$ being the
chain-direction displacement of the metal ($i=1,2$) or halogen
($i=3$) in the $n$th MXM unit from its equilibrium position.
$\alpha$ and $\beta$ are, respectively, the intersite and intrasite
el-ph coupling constants, while $K$ is the metal-halogen spring
constant.
We assume, based on the thus-far reported experimental observations,
that every M$_2$ moiety is not deformed, namely, $u_{1:n}=u_{2:n-1}$.
$\varepsilon_{\rm M}$ and $\varepsilon_{\rm X}$ are the on-site
energies (electron affinities) of isolated metal and halogen atoms,
respectively.
Since $\varepsilon_{\rm M}$ is usually larger than
$\varepsilon_{\rm X}$, we neglect the weak alternation of the
halogen-ion on-site energies.
The electron hoppings between these levels are modeled by
$t_{\rm MM}$ and $t_{\rm MX}$, whereas the electron-electron Coulomb
interactions by $U_{\rm M}$, $U_{\rm X}$, $V_{\rm MM}$, and
$V_{\rm MX}$.
Coulomb interactions between neighboring metal and halogen ions may
in principle alternate in accordance with lattice displacements in a
two-band description.
However, $V_{\rm MX}$ itself is rather small due to the ligands
surrounding metals and its alternate component $\delta V_{\rm MX}$ is
further reduced by the factor $l_{i:n}/r$ with $r$ being the M-X
distance in the undistorted lattice.
Therefore, any alternation of Coulomb interactions is usually
neglected.
We take $N$ for the number of unit cells in the following.
Let us rewrite the Hamiltonian compactly in the momentum space as
\widetext
\begin{eqnarray}
   {\cal H}
   &=& \sum_{i,j}\sum_{k,q}\sum_{s,s'}
       \langle i:k+q,s\vert t\vert j:k,s'\rangle
       a_{i:k+q,s}^\dagger a_{j:k,s'}
       \nonumber \\
   &+& \frac{1}{2}\sum_{i,j,m,n}\sum_{k,k',q}\sum_{s,s',t,t'}
       \langle i:k+q,s;m:k',t\vert v\vert j:k,s';n:k'+q,t'\rangle
       a_{i:k+q,s}^\dagger a_{m:k',t}^\dagger
       a_{n:k'+q,t'} a_{j:k,s'}
       \nonumber \\
   &+& \frac{K}{2}\sum_q
       \left[
        u_{1:q}u_{1:q}^*+u_{2:q}u_{2:q}^*+2u_{3:q}u_{3:q}^*
       -\big(
         {\rm e}^{ {\rm i}q/3}u_{1:q}u_{3:q}^*
        +{\rm e}^{-{\rm i}q/3}u_{2:q}u_{3:q}^*
        +{\rm c.c.}
        \big)
       \right]\,,
   \label{E:Hk}
\end{eqnarray}
\narrowtext
where
we have set the unit-cell length (the distance between neighboring
halogen ions) equal to unity.
The momentum representation of the interactions can straightforwardly
be obtained from the Fourier transform of the original Hamiltonian
(\ref{E:H}) and we learn
%\begin{eqnarray}
%   && \langle i:k+q,s\vert t\vert j:k,s'\rangle
%     =\langle i:k+q\vert t\vert j:k\rangle\delta_{ss'} \,,
%      \label{E:intt}
%      \\
%   && \langle i:k+q,s;m:k',t\vert v\vert j:k,s';n:k'+q,t'\rangle
%     =\langle i:k+q;m:k'\vert v\vert j:k;n:k'+q\rangle
%      \delta_{ss'}\delta_{tt'} \,.
%      \label{E:intv}
%\end{eqnarray}
\begin{eqnarray}
   && \langle i:k+q,s\vert t\vert j:k,s'\rangle
     =\langle i:k+q\vert t\vert j:k\rangle\delta_{ss'} \,,
      \label{E:intt}
      \\
   && \langle i:k+q,s;m:k',t\vert v\vert j:k,s';n:k'+q,t'\rangle
      \nonumber \\
   && \quad
     =\langle i:k+q;m:k'\vert v\vert j:k;n:k'+q\rangle
      \delta_{ss'}\delta_{tt'} \,.
      \label{E:intv}
\end{eqnarray}

   The most essential process of the Hartree-Fock approximation must
be the introduction of order parameters.
A symmetry argument \cite{Ozak83,Ozak14} allows us to systematically
derive density-wave solutions and reduce the following numerical
efforts to the minimum necessary.
As far as we consider the normal states, that is, unless the gauge
symmetry is broken, the symmetry group of the present system is given
by
\begin{equation}
   {\bf G}={\bf P}\times{\bf S}\times{\bf T} \,,
   \label{E:G}
\end{equation}
where ${\bf P}={\bf L}\land{\bf C}_2$ is the space group of a linear
chain with the one-dimensional translation group ${\bf L}$ whose
basis vector is the unit-cell translation $l$,
${\bf S}$ is the group of spin-rotation, and
${\bf T}$ is the group of time reversal.
Group actions on the electron operators are defined in Table
\ref{T:GA}, where
$l\in{\bf L}$,
$C_2\in{\bf C}_2$,
$u(\mbox{\boldmath$e$},\theta)
 =\sigma^0\cos(\theta/2)
 -(\mbox{\boldmath$\sigma$}\cdot\mbox{\boldmath$e$})
  \sin(\theta/2)
 \in{\bf S}$, and
$t\in{\bf T}$ with
$\sigma^0$ and
$\mbox{\boldmath$\sigma$}=(\sigma^x,\sigma^y,\sigma^z)$
being the $2\times 2$ unit matrix and a vector composed of the
Pauli-matrices, respectively.

   Let $\check{G}$ denote the irreducible representations of {\bf G}
over the real number field, where their representation space is
spanned by the Hermitian operators
\{$a_{i:k,s}^\dagger a_{j:k',s'}$\}.
There is a one-to-one correspondence between $\check{G}$ and
broken-symmetry phases of density-wave type \cite{Ozak55,Yama29}.
Any representation $\check{G}$ is obtained as a Kronecker product of
the irreducible real representations of ${\bf P}$, ${\bf S}$, and
${\bf T}$:
\begin{equation}
   \check{G}=\check{P}\otimes\check{S}\otimes\check{T} \,.
   \label{E:Grep}
\end{equation}
$\check{P}$ is characterized by an ordering vector $q$ in the
Brillouin zone and an irreducible representation of its little group
${\bf P}(q)$, and is therefore labeled $q\check{P}(q)$.
The relevant representations of ${\bf S}$ are given by
\begin{equation}
  \left.
  \begin{array}{ll}
   \check{S}^0(u(\mbox{\boldmath$e$},\theta))
     =1
   & \mbox{(nonmagnetic)} \,,\\
   \check{S}^1(u(\mbox{\boldmath$e$},\theta))
     =O(u(\mbox{\boldmath$e$},\theta))
   & \mbox{(magnetic)} \,,
  \end{array}
  \right.
  \label{E:Srep}
\end{equation}
where $O(u(\mbox{\boldmath$e$},\theta))$ is the $3\times 3$
orthogonal matrix satisfying
$
   u(\mbox{\boldmath$e$},\theta)
   \mbox{\boldmath$\sigma$}^\lambda
   u^\dagger(\mbox{\boldmath$e$},\theta)
      =\sum_{\mu=x,y,z}
       [O(u(\mbox{\boldmath$e$},\theta))]_{\lambda\mu}
       \mbox{\boldmath$\sigma$}^\mu \ \
       (\lambda=x,\,y,\,z)
$,
whereas those of ${\bf T}$ by
\begin{equation}
  \left.
  \begin{array}{ll}
    \check{T}^0(t)= 1 & \mbox{(symmetric)} \,,\\
    \check{T}^1(t)=-1 & \mbox{(antisymmetric)} \,.
  \end{array}
  \right.
  \label{E:Trep}
\end{equation}
The representations
$\check{P}\otimes\check{S}^0\otimes\check{T}^0$,
$\check{P}\otimes\check{S}^1\otimes\check{T}^1$,
$\check{P}\otimes\check{S}^0\otimes\check{T}^1$, and 
$\check{P}\otimes\check{S}^1\otimes\check{T}^0$
correspond to charge-density-wave (CDW), spin-density-wave (SDW),
charge-current-wave (CCW), and spin-current-wave (SCW) states,
respectively.
Here in one dimension, the current-wave states either result in
the one-way uniform flow or break the charge-conservation law, and
are therefore less interesting.
We consider density waves of $q=0$ and $q=\pi$, which are labeled
$\mit\Gamma$ and $X$, respectively.
Since ${\bf P}({\mit\Gamma})={\bf P}(X)={\bf C}_2$,
$\check{P}({\mit\Gamma})$ and $\check{P}(X)$ are either
$A$ ($C_2$-symmetric) or $B$ ($C_2$-antisymmetric) representation of
${\bf C}_2$.

\subsection{Broken-symmetry solutions}

   Now the mean-field Hamiltonian is given by
%\begin{equation}
%   {\cal H}_{\rm HF}
%      =\sum_{i,j}\sum_{k,s,s'}\sum_{\lambda}
%       \Bigl[
%        x_{ij}^{\lambda}({\mit\Gamma};k)
%        a_{i:k,s}^\dagger a_{j:k,s'}
%       +x_{ij}^{\lambda}(X;k)
%        a_{i:k+\pi,s}^\dagger a_{j:k,s'}
%       \Bigr]
%       \sigma_{ss'}^\lambda\,.
%   \label{E:HHF}
%\end{equation}
\begin{eqnarray}
   {\cal H}_{\rm HF}
     &=&\sum_{i,j}\sum_{k,s,s'}\sum_{\lambda}
        \Bigl[
         x_{ij}^{\lambda}({\mit\Gamma};k)
         a_{i:k,s}^\dagger a_{j:k,s'}
   \nonumber \\
     && \quad
        +x_{ij}^{\lambda}(X;k)
         a_{i:k+\pi,s}^\dagger a_{j:k,s'}
        \Bigr]
        \sigma_{ss'}^\lambda\,.
   \label{E:HHF}
\end{eqnarray}
Here the self-consistent fields
$x_{ij}^\lambda({\mit\Gamma};k)$ and $x_{ij}^\lambda(X;k)$
are described as
\begin{eqnarray}
   &&
   x_{ij}^0({\mit\Gamma};k)
     =\langle i:k|t|j:k \rangle
     +\sum_{m,n}\sum_{k'}
      \rho_{nm}^0({\mit\Gamma};k')
   \nonumber \\
   &&\qquad\quad\times
      \big(
       2\langle i:k;m:k'|v|j:k ;n:k' \rangle
   \nonumber \\
   &&\qquad\qquad
       -\langle i:k;m:k'|v|n:k';j:k  \rangle
      \big)\,,
   \\
   &&
   x_{ij}^0(X;k)
     =\langle i:k+\pi|t|j:k \rangle
     +\sum_{m,n}\sum_{k'}
      \rho_{nm}^0(X;k'+\pi)
   \nonumber \\
   &&\qquad\quad\times
      \big(
       2\langle i:k+\pi;m:k'|v|j:k     ;n:k'+\pi \rangle
   \nonumber \\
   &&\qquad\qquad
       -\langle i:k+\pi;m:k'|v|n:k'+\pi;j:k      \rangle
      \big)\,,
   \\
   &&
   x_{ij}^z({\mit\Gamma};k)
    =-\sum_{m,n}\sum_{k'}
      \rho_{nm}^z({\mit\Gamma};k')
   \nonumber \\
   &&\qquad\quad\times
      \langle i:k;m:k'|v|n:k';j:k \rangle\,,
   \\
   &&
   x_{ij}^z(X;k)
    =-\sum_{m,n}\sum_{k'}
      \rho_{nm}^z(X;k'+\pi)
   \nonumber \\
   &&\qquad\quad\times
      \langle i:k+\pi;m:k'|v|n:k'+\pi;j:k\rangle\,,
   \label{E:SCF}
\end{eqnarray}
%\begin{eqnarray}
%   &&
%  x_{ij}^0({\mit\Gamma};k)
%     =\langle i:k|t|j:k \rangle
%     +\sum_{m,n}\sum_{k'}
%      \rho_{nm}^0({\mit\Gamma};k')
%   \nonumber \\
%   &&\qquad\times
%      \big(
%       2\langle i:k;m:k'|v|j:k ;n:k' \rangle
%       -\langle i:k;m:k'|v|n:k';j:k  \rangle
%      \big)\,,
%   \\
%   &&
%   x_{ij}^0(X;k)
%     =\langle i:k+\pi|t|j:k \rangle
%     +\sum_{m,n}\sum_{k'}
%      \rho_{nm}^0(X;k'+\pi)
%   \nonumber \\
%   &&\qquad\times
%      \big(
%       2\langle i:k+\pi;m:k'|v|j:k     ;n:k'+\pi \rangle
%       -\langle i:k+\pi;m:k'|v|n:k'+\pi;j:k      \rangle
%      \big)\,,
%   \\
%   &&
%   x_{ij}^z({\mit\Gamma};k)
%    =-\sum_{m,n}\sum_{k'}
%      \rho_{nm}^z({\mit\Gamma};k')
%      \langle i:k;m:k'|v|n:k';j:k \rangle\,,
%   \\
%   &&
%   x_{ij}^z(X;k)
%    =-\sum_{m,n}\sum_{k'}
%      \rho_{nm}^z(X;k'+\pi)
%      \langle i:k+\pi;m:k'|v|n:k'+\pi;j:k\rangle\,,
%   \label{E:SCF}
%\end{eqnarray}
in terms of the density matrices
\begin{eqnarray}
   &&
   \rho_{ij}^\lambda({\mit\Gamma};k)
    =\frac{1}{2}\sum_{s,s'}
     \langle a_{j:k,s}^\dagger a_{i:k,s'}\rangle_{\rm HF}\,
     \sigma_{ss'}^\lambda \,,
   \label{E:rhoG}
   \\
   &&
   \rho_{ij}^\lambda(X;k)
    =\frac{1}{2}\sum_{s,s'}
     \langle a_{j:k+\pi,s}^\dagger a_{i:k,s'}\rangle_{\rm HF}\,
     \sigma_{ss'}^\lambda \,,
   \label{E:rhoM}
\end{eqnarray}
where $\langle\cdots\rangle_{\rm HF}$ denotes the quantum average in
a Hartree-Fock eigenstate.
We note that no helical SDW solution ($\lambda=x,y$) is obtained from
the present Hamiltonian.
We decompose the Hamiltonian (\ref{E:HHF}) as
\begin{equation}
  {\cal H}_{\rm HF}
   =\sum_{D=A,B}\sum_{K={\mit\Gamma},X}\sum_{\lambda=0,z}
    h_{KD}^\lambda\,.
\end{equation}
Here the irreducible component $h_{KD}^\lambda$ can be obtained
through a general formula
\begin{equation}
 h_{KD}^\lambda
 =\frac{d^{(D)}}{g}\sum_{p\in{\bf C}_2}
  \chi^{(D)}(p)\,p\cdot
  x_{ij}^\lambda(K;k)a_{i:k,s}^\dagger a_{j:k,s'}
  \sigma_{ss'}^\lambda\,,
\end{equation}
where $\chi^{(D)}(p)$ is the irreducible character of the $D$
representation for the group element $p$, $g(=2)$ is the order of
${\bf C}_2$, and $d^{(D)}(=1)$ is the dimension of $D$.
Thus we obtain the broken-symmetry Hamiltonian for the representation
$KD\otimes\check{S}^i\otimes\check{T}^i$ as
${\cal H}_{\rm HF}(K;D)=h_{{\mit\Gamma}A}^0+h_{KD}^\lambda$,
where $\lambda=0$ ($i=0$) or $\lambda=z$ ($i=1$).
We explicitly show $h_{KD}^\lambda$ in Appendix \ref{A:h}.

   In order to characterize each phase, we define local order
parameters.
The charge and spin densities on site $i$ at the $n$th MXM unit are,
respectively, given by
\begin{eqnarray}
   d_{i:n}
     &=&\sum_s\langle a_{i:n,s}^\dagger a_{i:n,s}\rangle_{\rm HF}
   \,, \label{E:OPd} \\
   s_{i:n}^z
     &=&\frac{1}{2}\sum_{s,s'}
     \langle a_{i:n,s}^\dagger a_{i:n,s'}\rangle_{\rm HF}
     \sigma_{ss'}^z
   \,, \label{E:OPs}
\end{eqnarray}
while the bond and spin bond orders between site $i$ at the $n$th MXM
unit and site $j$ at the $m$th MXM unit are, respectively, defined as
\begin{eqnarray}
   p_{i:n;j:m}
     &=&\sum_s
     \langle a_{i:n,s}^\dagger a_{j:m,s}\rangle_{\rm HF}
   \,, \label{E:OPp} \\
   t_{i:n;j:m}^z
     &=&\frac{1}{2}\sum_{s,s'}
     \langle a_{i:n,s}^\dagger a_{j:n,s'}\rangle_{\rm HF}
     \sigma_{ss'}^z
   \,. \label{E:OPt}
\end{eqnarray}
Though $p_{i:n;j:m}$ and $t_{i:n;j:m}^z$ can generally be complex,
their imaginary parts are necessarily zero in our argument.
The order parameters can also be described in terms of the density
matrices (\ref{E:rhoG}) and (\ref{E:rhoM}).
Demanding that the density matrices should have the same symmetry
properties as their {\it host Hamiltonian}, we can qualitatively
characterize all the density-wave solutions.
We briefly describe their properties in the following and explicitly
show the analytic consequences of the symmetry argument in
Appendix \ref{A:OP}.
\smallskip

(a) ${\mit\Gamma}A\otimes\check{S}^0\otimes\check{T}^0$:\ 
The paramagnetic state with the full symmetry {\bf G}, abbreviated as
PM.

(b) ${\mit\Gamma}B\otimes\check{S}^0\otimes\check{T}^0$:\ 
Bond order wave with polarized charge densities on the M$_2$
moieties, abbreviated as BOW.

(c) $XA           \otimes\check{S}^0\otimes\check{T}^0$:\ 
Charge density wave on the X sublattice with alternate polarized
charge densities on the M$_2$ moieties, abbreviated as X-CDW.

(d) $XB           \otimes\check{S}^0\otimes\check{T}^0$:\ 
Charge density wave on the M sublattice, abbreviated as M-CDW.

(e) ${\mit\Gamma}A\otimes\check{S}^1\otimes\check{T}^1$:\ 
Ferromagnetism on both M$_2$ and X sublattices, abbreviated as FM.

(f) ${\mit\Gamma}B\otimes\check{S}^1\otimes\check{T}^1$:\ 
Spin bond order wave with polarized spin densities on the M$_2$
moieties, abbreviated as SBOW.

(g) $XA           \otimes\check{S}^1\otimes\check{T}^1$:\ 
Spin density wave on the X sublattice with alternate polarized spin
densities on the M$_2$ moieties, abbreviated as X-SDW.

(h) $XB           \otimes\check{S}^1\otimes\check{T}^1$:\ 
Spin density wave on the M sublattice, abbreviated as M-SDW.
\smallskip

\begin{figure}
\begin{flushleft}
\qquad\quad\mbox{\psfig{figure=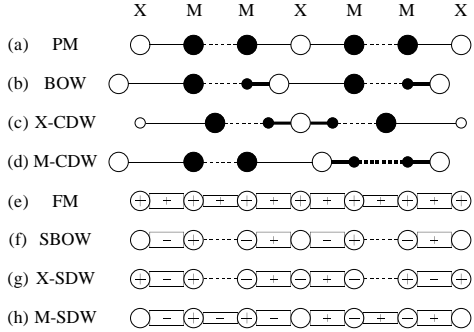,width=69mm,angle=0}}
\end{flushleft}
\vskip 2mm
\caption{Schematic representation of possible density-wave states,
         where the various circles and segments qualitatively
         represent the variation of local electron densities and
         bond orders, respectively, whereas the signs $\pm$ in
         circles and strips describe the alternation of local spin
         densities and spin bond orders, respectively.
         Circles shifted from the regular position qualitatively
         represent lattice distortion, which is peculiar to
         nonmagnetic phases.}
\label{F:DW}
\end{figure}

\noindent
Here,
${\bar\rho}_{ij}^\lambda(K;k)$ and ${\tilde\rho}_{ij}^\lambda(K;k)$
denote the real and imaginary parts of $\rho_{ij}^\lambda(K;k)$,
respectively.
All the phases are schematically shown in Fig. \ref{F:DW}.
The nonmagnetic broken-symmetry phases (a)-(d), respectively,
correspond to the oxidation states (a)-(d) in Fig. \ref{F:VS} in this
order.
%\vskip -6mm
\vskip 2mm

   So far little interest has been taken in the electronic state of
the halogen ions in the full belief that in comparison with the M
$d_{z^2}$ orbitals, the X $p_z$ orbitals are stably filled and less
effective, at least, in the ground state.
Such an idea may relatively be valid for MX chains, but it is not
the case any more for MMX chains where the metal sublattice can be
distorted.
Our symmetry argument suggests that the X-CDW (alternate
charge-polarization) state should be characterized by the charge
density wave on the X sublattice rather than any modulation on the
M$_2$ sublattice.
If we add up the charge and spin densities on the adjacent metal
sites in every M$_2$ moiety, they indeed modulate in M-CDW and
M-SDW, whereas they are uniform in X-CDW and X-SDW.
Alternate charge densities on the X sublattice necessarily mean the
oxidation of the halogen ions.

   We learn much more from the symmetry argument.
Magnetic instabilities are generally not coupled with phonons.
Specifying vanishing and nonvanishing density matrices and obtaining
the {\it symmetry-definite} Hartree-Fock energy expression, we find
which type of electronic correlation is effective in stabilizing each
density-wave state.
The halogen on-site Coulomb repulsion $U_{\rm X}$
stabilizes X-SDW and destabilizes X-CDW, respectively, while it is
much less relevant to the M-type density-wave states.
Though the intradimer different-site Coulomb repulsion $V_{\rm MM}$
has little effect on the X-type density-wave states, it efficiently
stabilizes both M-CDW and M-SDW states.
This is somewhat surprising because the repulsive interaction
$V_{\rm MM}$ favors charge disproportionation in the M$_2$ moiety at
the naivest consideration.
The $V_{\rm MM}$ stabilization of the M-type density-wave states
comes from the Hamiltonian component related to the field
$b_{XB}^\lambda$ (see Eqs. (\ref{E:hXB}), (\ref{E:SCFXB0}), and
(\ref{E:SCFXBz})).
We find that $V_{\rm MM}$ favorably contributes to the modulation of
the intradimer electron transfer.
The following numerical calculations show that the situation
$V_{\rm MM}>V_{\rm MX}$ generally stabilizes the M-type density-wave
states, whereas those of the X-type are relatively stabilized under
$V_{\rm MM}<V_{\rm MX}$, provided any other factor, such as the
alternation of transfer integrals, is put out of account.

\subsection{Ground-state phase diagrams}

\begin{figure}
\begin{flushleft}
\qquad\qquad\mbox{\psfig{figure=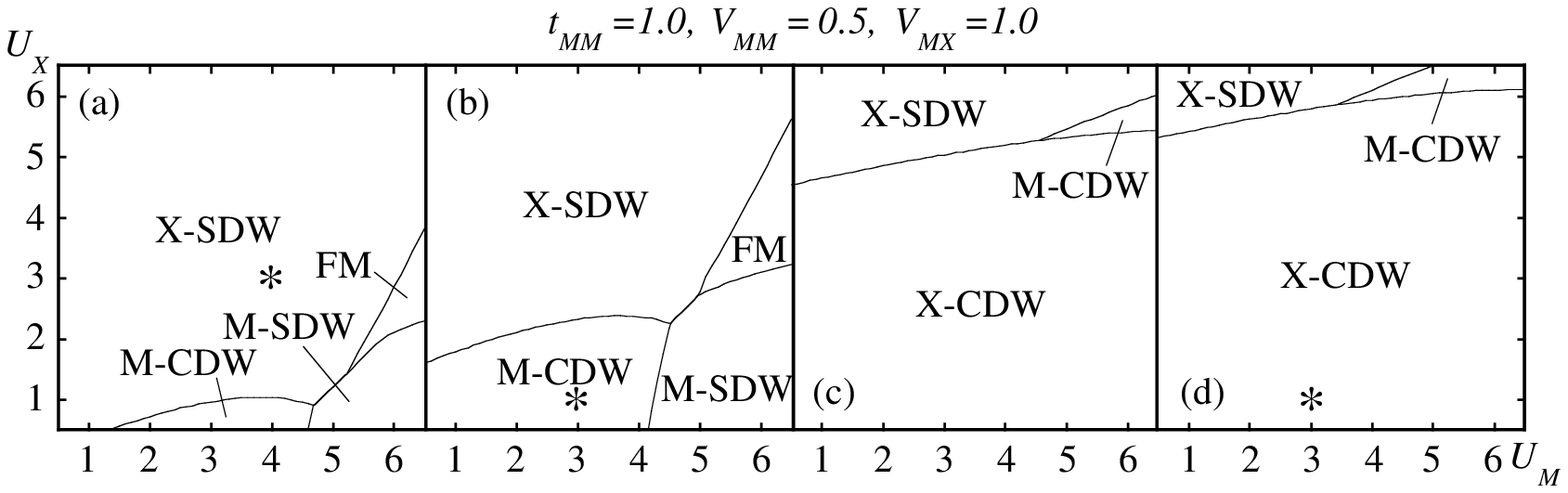,width=165mm,angle=0}}
\end{flushleft}
\begin{flushleft}
\qquad\qquad\mbox{\psfig{figure=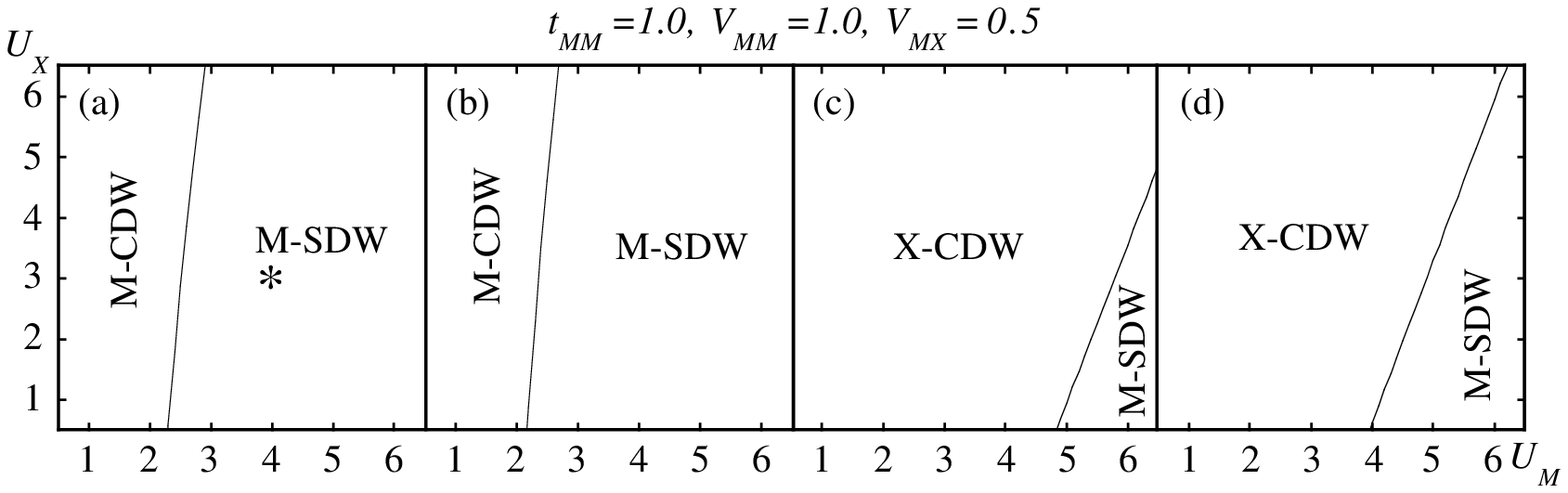,width=165mm,angle=0}}
\end{flushleft}
\begin{flushleft}
\qquad\qquad\mbox{\psfig{figure=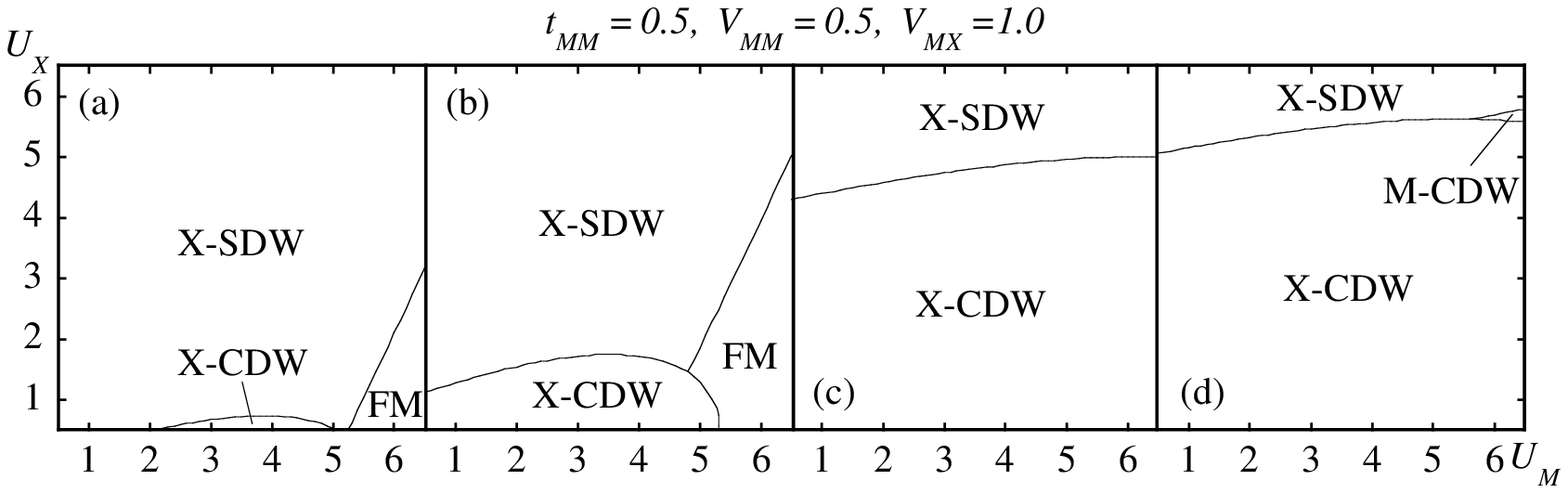,width=165mm,angle=0}}
\end{flushleft}
\begin{flushleft}
\qquad\qquad\mbox{\psfig{figure=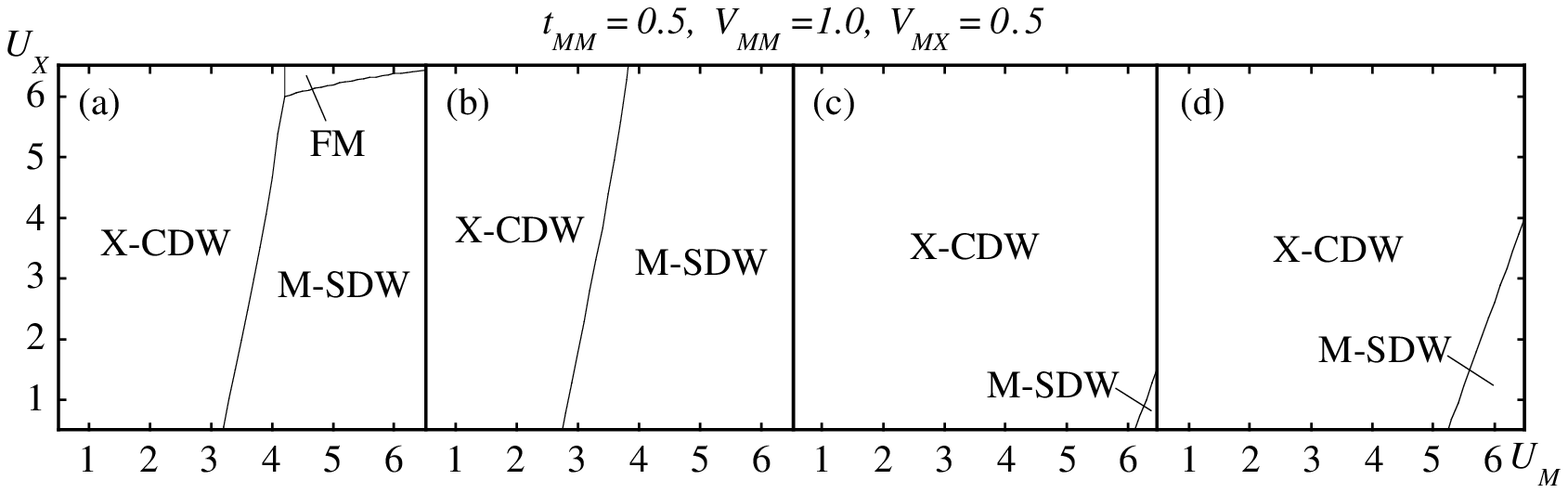,width=165mm,angle=0}}
\end{flushleft}
\vskip 5mm
\caption{Typical ground-state phase diagrams at $\frac{5}{6}$ band
         filling:
         (a) $\varepsilon_{\rm M}-\varepsilon_{\rm X}\equiv
              {\mit\Delta}\varepsilon=0.2,\alpha=0.6,\beta=1.8$;
         (b) ${\mit\Delta}\varepsilon=1.0,\alpha=0.6,\beta=1.8$;
         (c) ${\mit\Delta}\varepsilon=0.2,\alpha=1.8,\beta=0.6$;
         (d) ${\mit\Delta}\varepsilon=1.0,\alpha=1.8,\beta=0.6$.
         The rest of the parameters are common to each set of four
         figures and are indicated beside them.
         Asterisks in the phase diagrams are for the convenience of
         later arguments.}
\label{F:PhD}
\end{figure}

   Let us observe competing ground states numerically.
We calculate the Hartree-Fock energies
$\langle{\cal H}\rangle_{\rm HF}$ at a sufficiently low temperature
($k_{\rm B}T/t_{\rm MX}=0.025$) in the thermodynamic limit
($N\rightarrow\infty$).
We show in Fig. \ref{F:PhD} typical ground-state phase diagrams at
$\frac{5}{6}$ band filling, where the strength of any interaction is
indicated setting both $t_{\rm MX}$ and $K$ equal to unity.
In spite of plenty of parameters, the phase diagrams bring us a
general view of the ground-state properties:

  (i) Fundamentally, on-site Coulomb repulsions are advantageous to
      SDW states, whereas intersite repulsions to CDW states.
      This can be easily understood on the analogy of the
      {\it $U$-$V$ competition} between CDW and SDW states in the
      single-band extended Hubbard model.
      With increasing $U$, the antiferromagnetic spin alignments are
      replaced by FM.

 (ii) The site-diagonal electron-phonon coupling $\beta$ is favorable
      for M-CDW, while the intersite coupling $\alpha$ for X-CDW.
      This is also convincing when we simplify the argument employing
      a single-band model.
      Without the X $p_z$ orbitals, the dimerization of the X
      sublattice is simply modeled by modulating {\it on-site}
      electron affinities, while that of the M$_2$ sublattice results
      in the modulation of {\it intersite} hopping integrals just as
      polyacetylene \cite{Su98}.
      Thus, it is likely in the pop complexes that the interchain
      locking of the M$_2$ sublattices due to counter ions
      effectively realizes the parametrization $\beta>\alpha$.
      $\alpha$-versus-$\beta$ phase diagrams have recently been
      calculated within a single-band approximation by Kuwabara and
      Yonemitsu \cite{Kuwa}, where the competition from this point of
      view can be observed in more detail.
      They report that $\beta$ and $\alpha$ favor M-CDW and X-CDW,
      respectively, but X-CDW is much more widely stabilized to the
      ground state than M-CDW, which is consistent with the present
      observations.

\widetext
\narrowtext

(iii) As far as the Coulomb interactions are concerned, $U_{\rm M}$
      and $V_{\rm MM}$ contribute to the M-type density waves
      appearing, while $U_{\rm X}$ and $V_{\rm MX}$ to those of the
      X-type.
      If we neglect the intersite Coulomb repulsions and any
      modulation of transfer integrals, the phase diagram is rather
      understandable:
      For small $U$, M-CDW ($\alpha<\beta$) or X-CDW
      ($\alpha\agt\beta$); for large $U$, FM; otherwise, M-SDW
      ($U_{\rm M}\agt U_{\rm X}-{\mit\Delta}\varepsilon$) or X-SDW
      ($U_{\rm M}\alt U_{\rm X}-{\mit\Delta}\varepsilon$).
      ${\mit\Delta}\varepsilon
       \equiv\varepsilon_{\rm M}-\varepsilon_{\rm X}$ may be
      recognized as the orbital-energy correction to the
      {\it M-X competition} \cite{Yama13}.

 (iv) The competition between M-CDW and X-CDW is sensitive to the
      modulation of transfer integrals as well.
      Provided $\alpha$ is not so large as $\beta$, M-CDW is
      stabilized under $t_{\rm MM}>t_{\rm MX}$, whereas it is
      replaced by X-CDW under $t_{\rm MM}\alt t_{\rm MX}$.
      The orbital hybridization within every M$_2$ moiety is
      essential in the {\it valence-trapped} M-CDW state, while it is
      the overlap of the $d_{\sigma^*}$ orbitals on neighboring M$_2$
      moieties that stabilizes the {\it valence-delocalized} X-CDW
      state.
      Based on semiempirical quantum-chemical band calculations,
      Borshch {\it et al.} \cite{Bors62} pointed out a possibility of
      the X-CDW-type valence structure being stabilized in this
      context.
      They revealed, restricting their argument to the
      Pt$_2$(dta)$_4$I compound, that twisting of the dta ligand
      reduces the electronic communication by $\pi$ delocalization
      between adjacent metal sites and ends in charge
      disproportionation within the M$_2$ moiety.
      On the other hand, employing a single-band model, Baeriswyl and
      Bishop \cite{Baer39} pioneeringly suggested that with
      increasing M-X-M interdimer transfer integral, the distortion
      of the halogen sublattice should shrink and in the end the
      metal sublattice may begin to dimerize.
      Though their argument was mainly devoted to the MX-chain
      system, it is valid enough for the MMX-chain system as well.
      Such a behavior was indeed observed \cite{Swan05} applying
      pressure to a MMX material
      K$_4$[Pt$_2$(pop)$_4$Br]$\cdot$3H$_2$O, which may be understood
      as a pressure-induced reverse Peierls instability.
\smallskip

   At $\frac{5}{6}$ band filling, due to the perfect nesting of the
Fermi surface with respect to $q=\pi$ phonons, $X(q=\pi)$-phases are
predominantly stabilized.
Therefore, doping of the system generally contributes to the
stabilization of ${\mit\Gamma}(q=0)$-phases, though any doping of
MMX chains has not yet been reported.
At $\frac{4}{6}$ band filling, for example, an interesting
competition between BOW and SBOW \cite{Yama83} is observed, that is,
the competition between {\it site-centered} CDW and SDW states is
replaced by that between {\it bond-centered} CDW and SDW states.

   Now we are convinced that a single-band description of the M-type
density waves (M-CDW and M-SDW) may be justified well, whereas
the halogen $p_z$ orbitals should explicitly be taken into
consideration in describing those of the X-type (X-CDW and X-SDW).
In the next section, we further support this criterion using a
quantum Monte Carlo technique.
The pop and dta families will fully be characterized as approximate
$d_{z^2}$-single-band materials and really $d$-$p$-hybridized two-band
materials, respectively.

\begin{figure}
\begin{flushleft}
\quad\ \mbox{\psfig{figure=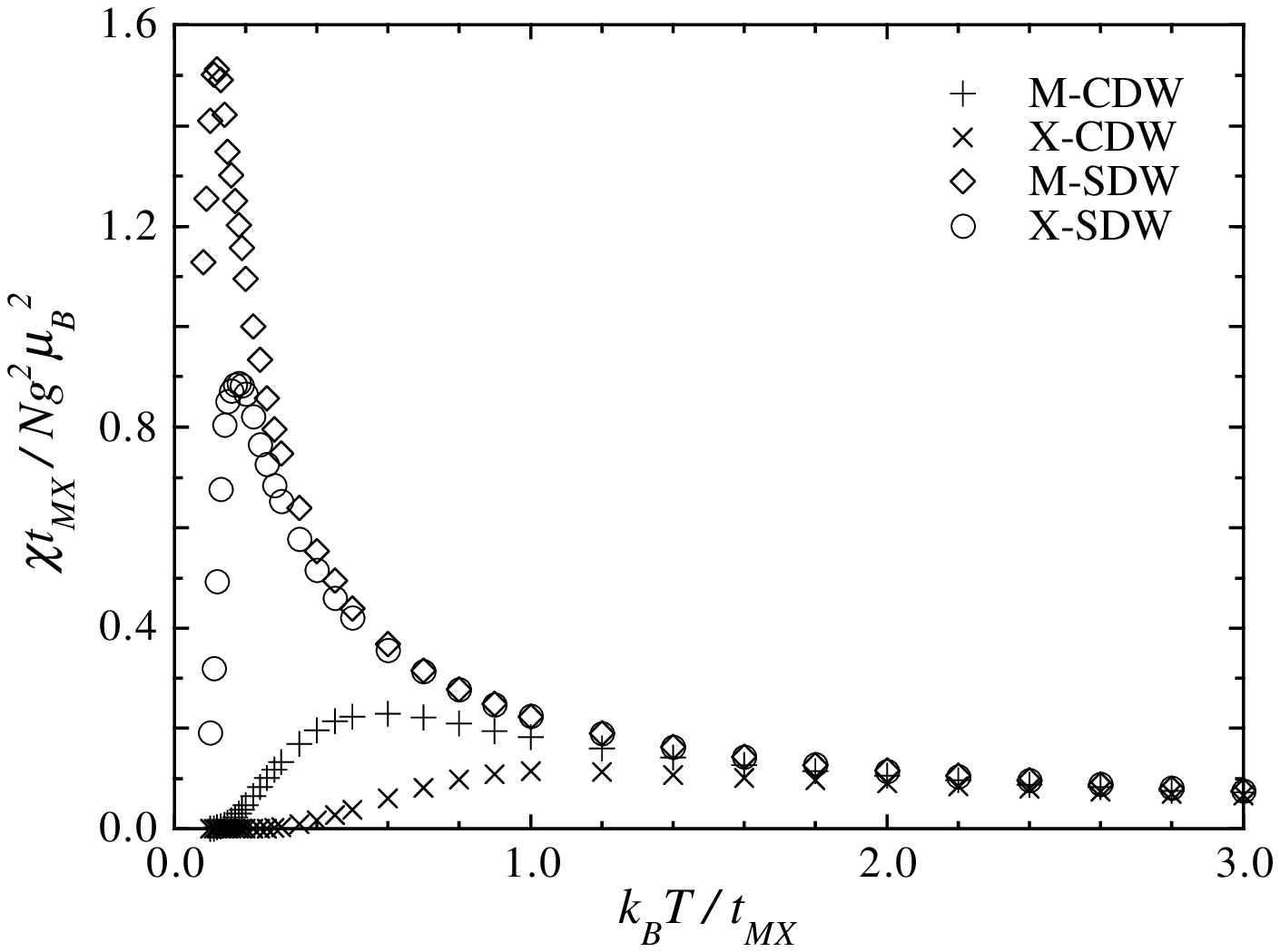,width=70mm,angle=0}}
\end{flushleft}
\vskip -2mm
\caption{Temperature dependences of the magnetic susceptibility at
         typical parametrizations:
         $t_{\rm MM}=1.0$, ${\mit\Delta}\varepsilon=1.0$,
          $\alpha=0.6$, $\beta=1.8$,
          $U_{\rm M}=3.0$, $U_{\rm X}=1.0$,
          $V_{\rm MM}=0.5$, and $V_{\rm MX}=1.0$ (M-CDW);
         $t_{\rm MM}=1.0$, ${\mit\Delta}\varepsilon=1.0$,
          $\alpha=1.8$, $\beta=0.6$,
          $U_{\rm M}=3.0$, $U_{\rm X}=1.0$,
          $V_{\rm MM}=0.5$, and $V_{\rm MX}=1.0$ (X-CDW);
         $t_{\rm MM}=1.0$, ${\mit\Delta}\varepsilon=0.2$,
          $\alpha=0.6$, $\beta=1.8$,
          $U_{\rm M}=4.0$, $U_{\rm X}=3.0$,
          $V_{\rm MM}=1.0$, and $V_{\rm MX}=0.5$ (M-SDW); and
         $t_{\rm MM}=1.0$, ${\mit\Delta}\varepsilon=0.2$,
          $\alpha=0.6$, $\beta=1.8$,
          $U_{\rm M}=4.0$, $U_{\rm X}=3.0$,
          $V_{\rm MM}=0.5$, and $V_{\rm MX}=1.0$ (X-SDW),
         where the M-CDW-, X-CDW-, M-SDW-, and X-SDW-type ground
         states are, respectively, obtained within the
         Hartree-Fock approximation.
         The four parametrizations are indicated by asterisks in
         Fig. 3.}
\label{F:chi}
\vskip 6mm
\begin{flushleft}
\quad\,\mbox{\psfig{figure=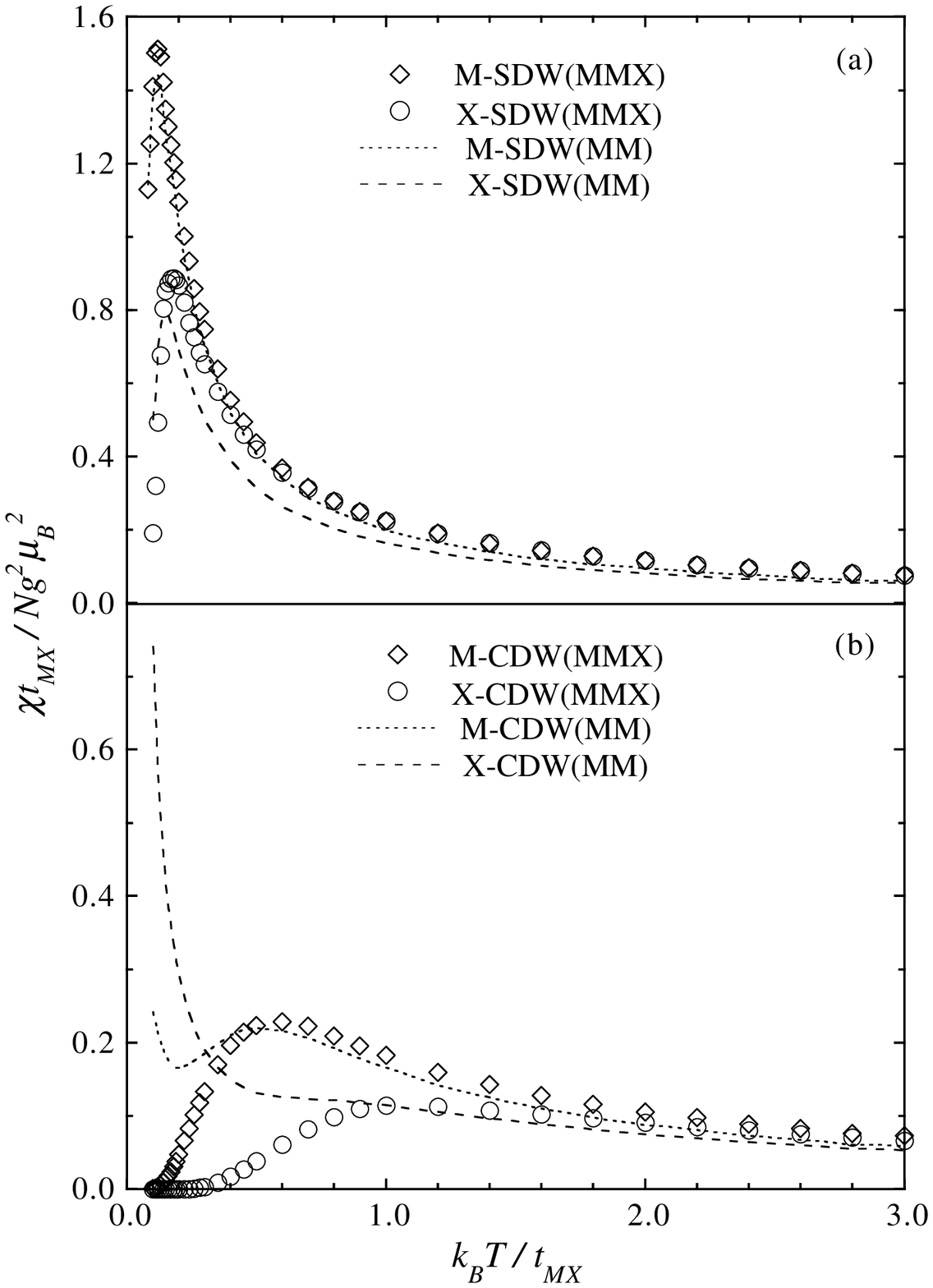,width=82mm,angle=0}}
\end{flushleft}
\vskip -2mm
\caption{Temperature dependences of the magnetic susceptibility
         (dotted and broken lines) on the assumption that the
         electrons occupying the X $p_z$ orbitals make no
         contribution to the magnetization, together with those
         (open diamonds and circles) coming from all the electrons.
         Parametrizations are the same as those in Fig. 4.}
\label{F:chiMM}
\end{figure}

\section{Quantum Monte Carlo approach}

\subsection{Numerical procedure}

   We employ a quantum Monte Carlo method \cite{Suzu54} based on the
checkerboard-type decomposition \cite{Hirs28} of the partition
function.
We simulate the Hamiltonian (\ref{E:H}), namely, we here make an
adiabatic description of the phonon field.
Starting with an arbitrary electron and lattice configuration, we
update the electron configuration under the frozen phonon field and
then update the lattice configuration under the fixed electron
configuration by the heat-bath algorithm.
At each Monte Carlo step, under the constraint $u_{1:n}=u_{2:n-1}$,
every site moves right or left by a sufficiently small unit
${\mit\Delta}_{\rm unit}$ or otherwise stays at the present position:
${\tilde u}_{i:n}\rightarrow{\tilde u}_{i:n}+\delta{\tilde u}_{i:n}$;
$\delta{\tilde u}_{i:n}
 ={\mit\Delta}_{\rm unit},-{\mit\Delta}_{\rm unit},0$,
where ${\tilde u}_{i:n}=\alpha u_{i:n}/t_{\rm MX}$.
${\mit\Delta}_{\rm unit}$ is set equal to $0.02$.
According to such movements of lattice sites, the bond configuration
is updated as
\begin{eqnarray}
  &&
  \left\{
  \begin{array}{lll}
   l_{1:n  } & \rightarrow & l_{1:n  }-\delta u_{1:n} \\
   l_{2:n-1} & \rightarrow & l_{2:n-1}-\delta u_{1:n}
  \end{array}
  \right.\ \,
  {\rm for}\ \,u_{1:n}\rightarrow u_{1:n}+\delta u_{1:n}\,,
  \nonumber \\
  &&
  \left\{
  \begin{array}{lll}
   l_{1:n} & \rightarrow & l_{1:n}+\delta u_{3:n} \\
   l_{2:n} & \rightarrow & l_{2:n}+\delta u_{3:n}
  \end{array}
  \right.\ \,
  {\rm for}\ \,u_{3:n}\rightarrow u_{3:n}+\delta u_{3:n}\,,
  \nonumber
\end{eqnarray}
where the total length of the lattice is kept unchanged.

   Since it is difficult to handle negative weights with the
world-line Monte Carlo algorithm \cite{Suzu54}, we further modify the
original model (\ref{E:H}) taking the metal-halogen hopping matrix
elements to be
\begin{eqnarray}
  &&
  \langle 1:n|t|3:n\rangle
   =\left\{
    \begin{array}{l}
     t_{\rm MX}-\alpha l_{1:n} \\
     \quad{\rm for}\ \,{\tilde u}_{3:n}-{\tilde u}_{1:n}<    1\,,\\
     0\ \,{\rm for}\ \,{\tilde u}_{3:n}-{\tilde u}_{1:n}\geq 1\,,
    \end{array}
    \right. \\
  &&
  \langle 2:n|t|3:n\rangle
   =\left\{
    \begin{array}{l}
     t_{\rm MX}+\alpha l_{2:n} \\
     \quad{\rm for}\ \, {\tilde u}_{2:n}-{\tilde u}_{3:n}<    1\,,\\
     0\ \,{\rm for}\ \, {\tilde u}_{2:n}-{\tilde u}_{3:n}\geq 1\,.
    \end{array}
    \right.
\end{eqnarray}
%\begin{eqnarray}
%  &&
%  \langle 1:n|t|3:n\rangle
%   =\left\{
%    \begin{array}{l}
%     t_{\rm MX}-\alpha l_{1:n}\ \,
%          {\rm for}\ \,{\tilde u}_{3:n}-{\tilde u}_{1:n}<    1\,,\\
%     0\ \,{\rm for}\ \,{\tilde u}_{3:n}-{\tilde u}_{1:n}\geq 1\,,
%    \end{array}
%    \right. \\
%  &&
%  \langle 2:n|t|3:n\rangle
%   =\left\{
%    \begin{array}{l}
%     t_{\rm MX}+\alpha l_{2:n}\ \,
%          {\rm for}\ \, {\tilde u}_{2:n}-{\tilde u}_{3:n}<    1\,,\\
%     0\ \,{\rm for}\ \, {\tilde u}_{2:n}-{\tilde u}_{3:n}\geq 1\,.
%    \end{array}
%    \right.
%\end{eqnarray}
As the atoms move away from each other, the overlap matrix elements
go to zero and do not change sign.
Therefore, the above modification makes sense from a physical point
of view and may be justified with not-so-large coupling constants.

\subsection{Magnetic susceptibility}

   We show in Fig. \ref{F:chi} temperature dependences of the
magnetic susceptibility $\chi$ peculiar to M-CDW, X-CDW, M-SDW, and
X-SDW.
At every temperature, we have taken several Monte Carlo estimates
changing the Trotter number $n_{\rm T}$ and their $n_{\rm T}$
dependence has been extrapolated to the $n_{\rm T}\rightarrow\infty$
limit.
With decreasing temperature, the susceptibility shows a gentle upward
slope and then exponentially vanishes in the CDW region, whereas in
the SDW region it exhibits a significant enhancement, which is
reminiscent of the low-temperature diverging behavior in Ni complexes
\cite{Maru99}.
The two CDW-proper susceptibilities show differences as well as
similarities.
The much more rapid low-temperature decrease in the X-CDW state is
likely to denote spin-Peierls like fluctuations.
It is convincing that the susceptibility in the
electronic-correlation-dominant region should also be saturated and
turn decreasing at temperatures which are low enough to develop
antiferromagnetic interactions between induced spin moments.
However, the Monte Carlo estimates of the SDW-proper behavior at low
temperatures are less convergent, where the numerical uncertainty is
more than the symbol size.
We should fundamentally take special care how we understand the
present SDW-proper susceptibilities, because no antiferromagnetic
spin-density long-range order is stabilized purely in one dimension.
If we calculate beyond the mean-field approximation, any SDW gap is
not obtained but SDW-like fluctuations grow more and more with
decreasing temperature in the relevant region.
However, the slightest interchain coupling stabilizes such
fluctuations into a long-range order.
The SDW states \cite{Bell15,Okam38} widely observed in Ni complexes
are thus stabilized, where hydrogen bonds and/or van der Waals
contacts supply weak but essential interchain interactions.
Therefore, concerning the SDW states as well, the present
calculations remain suggestive.
We learn that induced local spin moments are larger in the M-SDW
state than in the X-SDW state.
%This is understandable considering that spin moments are mainly
%induced on the X sites in the X-SDW state, whereas all the adjacent
%M sites cooperatively contribute electrons to spin moments in the
%M-SDW state.

   In order to clarify the contribution from the X $p_z$ orbitals, we
carry out further calculations of the susceptibility separating
magnetizations on the M$_2$ and X sublattices.
We define the modified magnetic susceptibility $\chi^{(n_{\rm orb})}$
as
\begin{equation}
   \chi^{(n_{\rm orb})}
    =\frac{g^2\mu_{\rm B}^2}{k_{\rm B}T}
     \Bigl[
      \big\langle(M^{(n_{\rm orb})})^2\big\rangle_{\rm th}
     -\big\langle M^{(n_{\rm orb})}\big\rangle_{\rm th}^2
     \Bigr]\,,
\end{equation}
where $\langle\cdots\rangle_{\rm th}$ denotes the thermal average at
a given temperature and
\begin{equation}
   M^{(n_{\rm orb})}
    =\sum_{n=1}^N\sum_{i=1}^{n_{\rm orb}}\sum_{s=\pm}
     \frac{s}{2}n_{i:n,s}\,.
\end{equation}
The susceptibility coming from the M $d_{z^2}$ orbitals,
$\chi^{\rm MM}$, is given by $\chi^{(2)}$, while $\chi^{(3)}$ is the
total susceptibility $\chi^{\rm MMX}$, which has just shown in Fig.
\ref{F:chi}.
We compare $\chi^{\rm MM}$ with $\chi^{\rm MMX}$ in Fig.
\ref{F:chiMM}.
We find a clear contrast between $\chi^{\rm MMX}$ and $\chi^{\rm MM}$
in the parameter region stabilizing X-CDW to the ground state.
Spin moments almost come from the M$_2$ sublattice at high
temperatures, where the system is expected to lie in the
averaged-valence state, while $\chi^{\rm MM}$ more and more deviates
from $\chi^{\rm MMX}$ and exhibits a diverging behavior with
decreasing temperature.
Thus, the X-CDW state is never stabilized without the
$p$-{\it electrons} compensating the spin moments on the M$_2$
sublattice.
The M-CDW state also exhibits such an aspect but it is much less
remarkable.
There exists a similar contrast between M-SDW and X-SDW as well.
The halo-
\begin{figure}
\begin{flushleft}
\quad\mbox{\psfig{figure=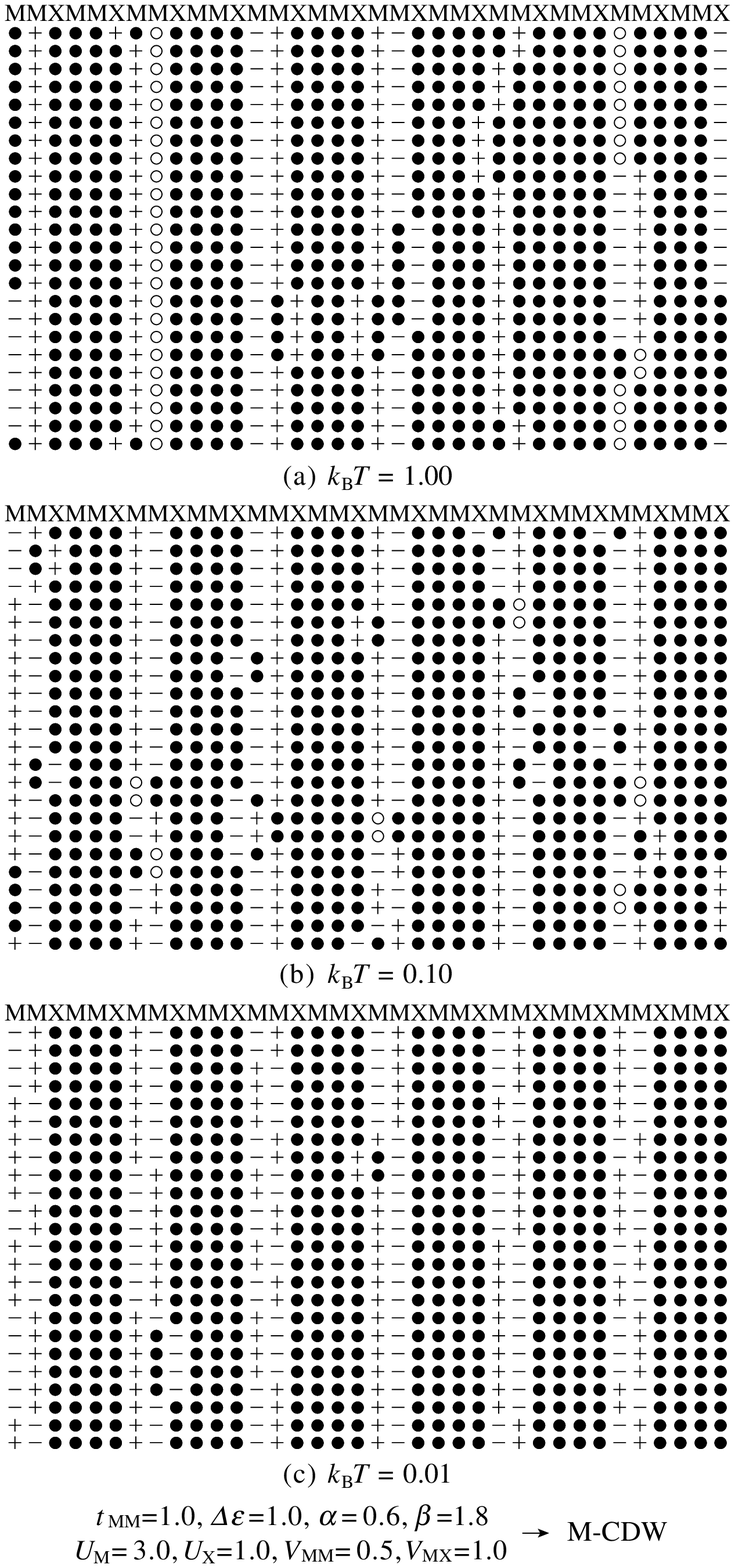,width=78mm,angle=0}}
\end{flushleft}
\vskip 0mm
\caption{Quantum Monte Carlo snapshots of the transformed
         two-dimensional Ising system of $N=12$, where the horizontal
         and the vertical axes correspond to space (the chain
         direction) and time (the Trotter direction), respectively.
         Trotter number $n_{\rm T}$ has been set equal to $12$, $24$,
         and $48$, and the whole time passage ($1/k_{\rm B}T$), the
         former half of that ($1/2k_{\rm B}T$), and the first quarter
         of that ($1/4k_{\rm B}T$) are shown for $k_{\rm B}T=1.00$,
         $0.10$, and $0.01$, respectively.
         $+$ and $-$ denote up and down spins, whereas $\circ$ and
         $\bullet$ represent vacant and doubly occupied sites,
         respectively.
         The parametrization is
         $t_{\rm MM}=1.0$, ${\mit\Delta}\varepsilon=1.0$,
         $\alpha=0.6$, $\beta=1.8$,
         $U_{\rm M}=3.0$, $U_{\rm X}=1.0$,
         $V_{\rm MM}=0.5$, and $V_{\rm MX}=1.0$,
         which stabilizes the M-CDW-type ground state.}
\label{F:snapMCDW}
\end{figure}
\vskip 0mm
\noindent
gen ions contribute no effective electron to the susceptibility
$\,$in$\,$ the$\,$ M-SDW$\,$ state,$\,$ whereas$\,$ all$\,$ the$\,$
electrons
\begin{figure}
\begin{flushleft}
\quad\mbox{\psfig{figure=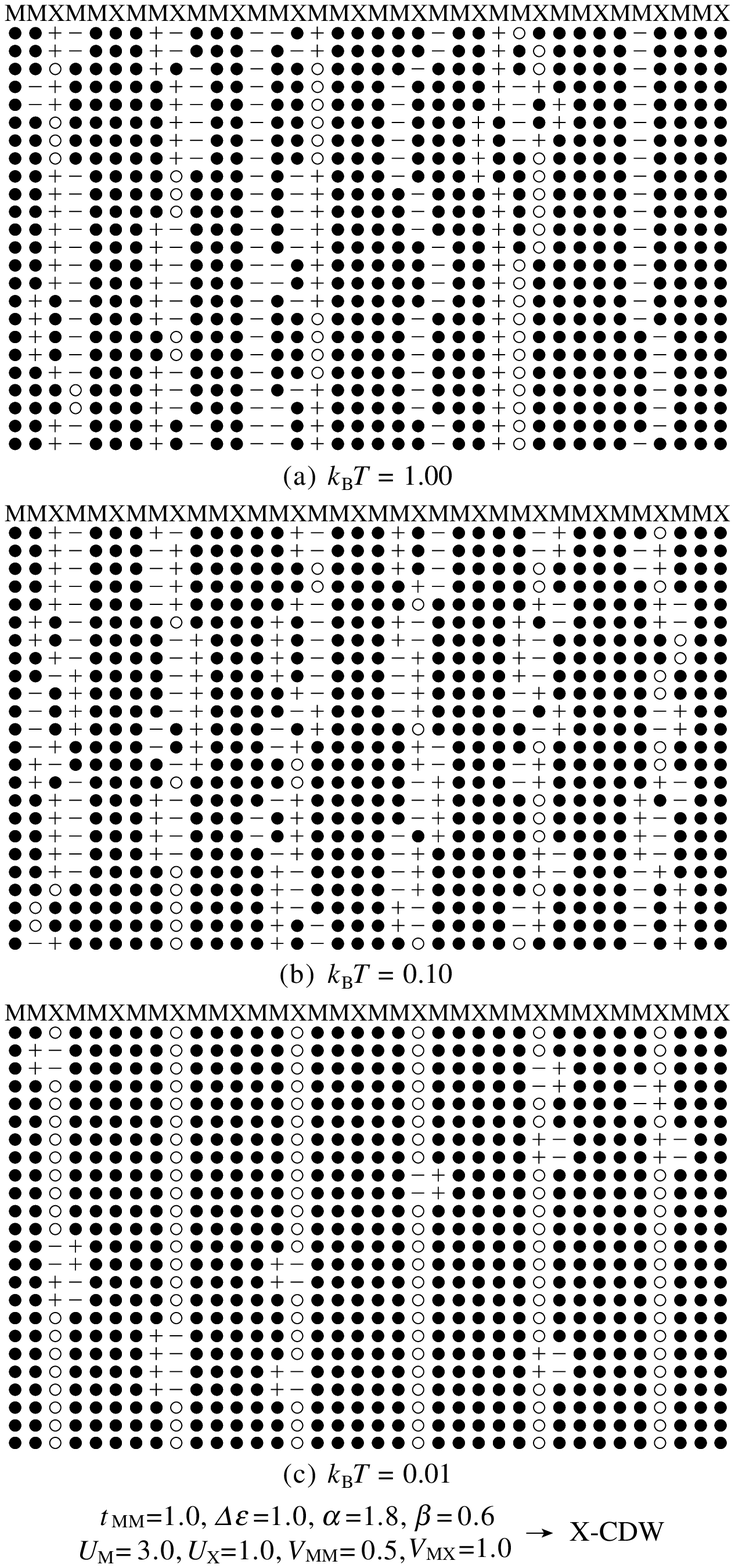,width=78mm,angle=0}}
\end{flushleft}
\vskip -1mm
\caption{The same as Fig. 6 but
         $t_{\rm MM}=1.0$, ${\mit\Delta}\varepsilon=1.0$,
         $\alpha=1.8$, $\beta=0.6$,
         $U_{\rm M}=3.0$, $U_{\rm X}=1.0$,
         $V_{\rm MM}=0.5$, and $V_{\rm MX}=1.0$,
         which stabilizes the X-CDW-type ground state.}
\label{F:snapXCDW}
\end{figure}
\vskip 0mm
\noindent
are effective in the X-SDW state.
We stress again that the contribution from the X sublattice,
$\chi^{\rm MMX}-\chi^{\rm MM}$, is not small in the X-SDW state,
noting the different scales in Figs. \ref{F:chiMM}(a) and
\ref{F:chiMM}(b).

\subsection{Direct observation of the electronic states}

\begin{figure}
\begin{flushleft}
\quad\mbox{\psfig{figure=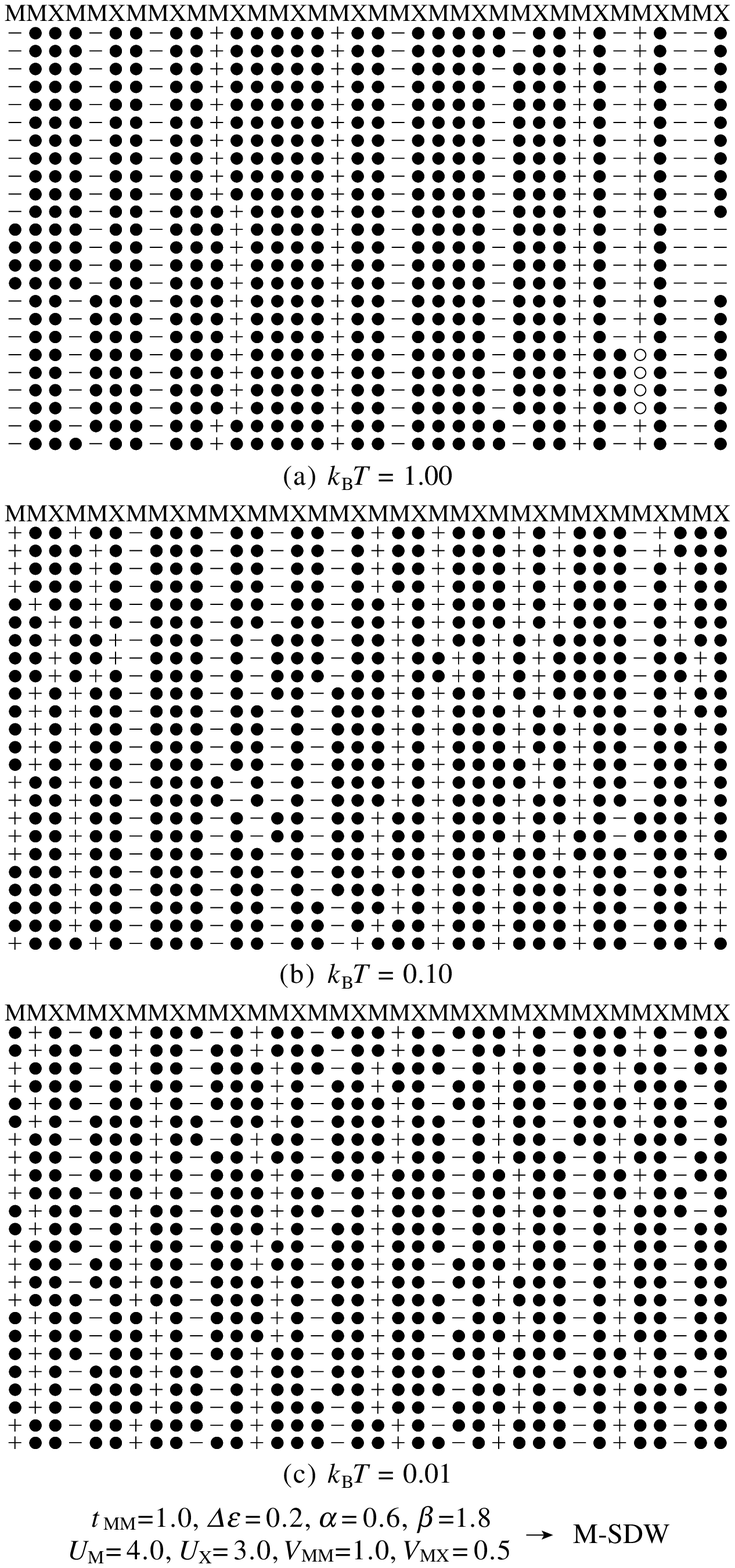,width=78mm,angle=0}}
\end{flushleft}
\vskip 0mm
\caption{The same as Fig. 6 but
         $t_{\rm MM}=1.0$, ${\mit\Delta}\varepsilon=0.2$,
         $\alpha=0.6$, $\beta=1.8$,
         $U_{\rm M}=4.0$, $U_{\rm X}=3.0$,
         $V_{\rm MM}=1.0$, and $V_{\rm MX}=0.5$,
         which is expected to be favorable for the M-SDW-type
         ground state.}
\label{F:snapMSDW}
\end{figure}

\begin{figure}
\begin{flushleft}
\quad\mbox{\psfig{figure=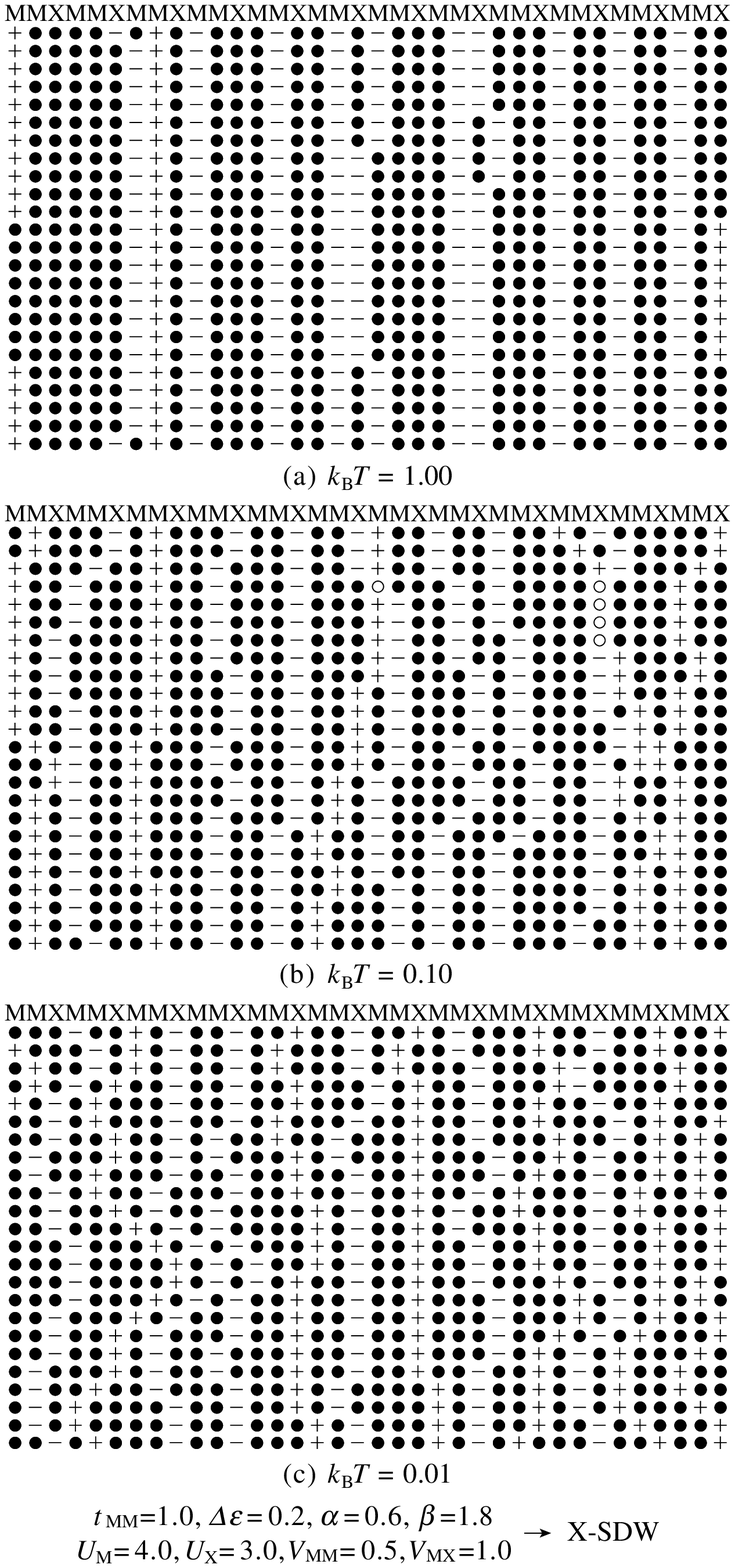,width=78mm,angle=0}}
\end{flushleft}
\vskip -1mm
\caption{The same as Fig. 6 but
         $t_{\rm MM}=1.0$, ${\mit\Delta}\varepsilon=0.2$,
         $\alpha=0.6$, $\beta=1.8$,
         $U_{\rm M}=4.0$, $U_{\rm X}=3.0$,
         $V_{\rm MM}=0.5$, and $V_{\rm MX}=1.0$,
         which is expected to be favorable for the X-SDW-type
         ground state.}
\label{F:snapXSDW}
\vskip 1mm
\end{figure}

   Any path-integral method has the advantage of directly visualizing
electronic states.
We show in Figs. \ref{F:snapMCDW}-\ref{F:snapXSDW} Monte Carlo
snapshots taken under the same parametri-
\widetext
\narrowtext
\widetext
\vskip 0.5mm
\narrowtext
\noindent
zations as the susceptibility
calculations.
Figures \ref{F:snapMCDW} and \ref{F:snapXCDW} clearly reveal the
roles played by the $d$ and $p$ orbitals in the CDW states.
The characteristic charge orders grow with decreasing temperature.
In the ground state of the M-CDW type, the X $p_z$ orbitals are
almost completely filled, whereas oxidized halogen ions appear in
every other unit in the X-CDW-type ground state.
It is obvious that the alternate charge densities on the X
sublattice, rather than the alternation of polarized charge densities
on the M$_2$ sublattice, characterize the X-CDW state.
Figures \ref{F:snapMSDW} and \ref{F:snapXSDW} also give a useful
piece of information about the SDW states.
The symmetry argument has concluded as follows:
The antiferromagnetic spin ordering on the X sublattice characterizes
X-SDW, where the M $d_{z^2}$ orbitals also contribute spin moments
favoring a parallel alignment within each MXM unit
(Fig. \ref{F:DW}(g));
The antiferromagnetic spin ordering on the M$_2$ sublattice
characterizes M-SDW, where the spin moments favor a parallel
alignment within each M$_2$ moiety (Fig. \ref{F:DW}(h)).
Figures \ref{F:snapMSDW} and \ref{F:snapXSDW} convincingly display
all these properties.
We stress in particular that the X $p_z$ orbitals are all filled in
Fig. \ref{F:snapMSDW}(c), as well as in Fig. \ref{F:snapMCDW}(c),
which well shows the single-band character of the M-type density-wave
states.
Here we should be reminded that the present observations come from
purely one-dimensional finite-chain calculations where any true
antiferromagnetic {\it long-range} order can neither be stabilized
nor be verified.
The spin configuration indeed deviates from the perfect order
peculiar to X-SDW here and there in Fig. \ref{F:snapXSDW}(c).
We find that domain walls prevent the characteristic local spin
alignments from growing into the long-range order.
Figure \ref{F:snapMSDW}(c) just happens to contain no domain wall
but longer-chain calculations still show the breakdown of the
long-range order.
It is, however, likely that M-SDW possesses more pronounced
low-dimensional character than X-SDW.
\vskip -2mm
\begin{figure}
\begin{flushleft}
\mbox{\psfig{figure=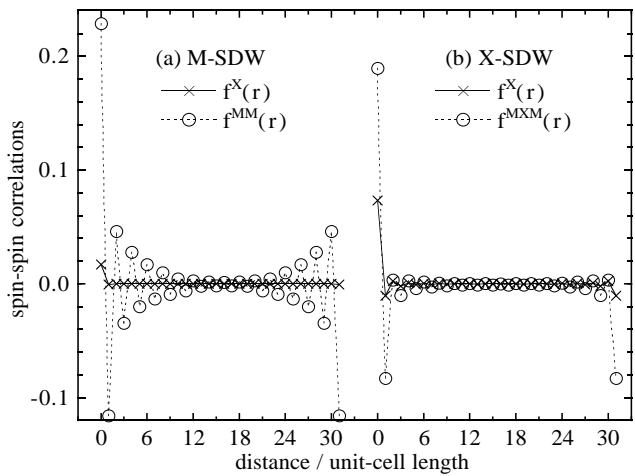,width=100mm,angle=0}}
\end{flushleft}
\vskip 1mm
\caption{Ground-state correlations between $r$-distant local spin
         moments on the X sites ($f^{\rm X}(r)$), the M$_2$ moieties
         ($f^{\rm MM}(r)$), and the MXM units ($f^{\rm MXM}(r)$) in
         the $N=32$ chain:
         (a) $f^{\rm X}(r)$ and $f^{\rm MM} (r)$ are shown by
         $\times$ and $\circ$, respectively, under the same
         parametrization as Fig. 8;
         (b) $f^{\rm X}(r)$ and $f^{\rm MXM}(r)$ are shown by
         $\times$ and $\circ$, respectively, under the same
         parametrization as Fig. 9.
         The solid and dotted lines are guides for eyes.
         It turns out that unless the Trotter number is large enough,
         the Monte Carlo estimates of ground-state spin-spin
         correlations for the present model are not so convergent but
         tend to be nonvanishingly frozen up.
         Therefore, setting $k_{\rm B}T$ equal to $0.01$, we thus
         present the conclusive calculations at a sufficiently large
         Trotter number, $n_{\rm T}=200$, rather than extrapolate
         less convergent estimates at smaller Trotter numbers.}
\label{F:SC}
\end{figure}
\vskip 1mm

   It is also helpful in characterizing the SDW states to observe
spin-spin correlations.
We show in Fig. \ref{F:SC} spin correlation functions of a few types
which are defined as
\begin{eqnarray}
   &&
   f^{\rm X}(r)
    = \sum_{s,s'=\pm}\frac{ss'}{4}
      \langle
       n_{3:n,s}n_{3:n+r,s'}
      \rangle_{\rm GS}\,,
   \nonumber \\
   &&
   f^{\rm MM}(r)
    = \sum_{i,j=1}^2\sum_{s,s'=\pm}\frac{ss'}{4}
      \langle
       n_{i:n,s}n_{j:n+r,s'}
      \rangle_{\rm GS}\,,
   \label{E:SC} \\
   &&
   f^{\rm MXM}(r)
    = \sum_{i,j=1}^3\sum_{s,s'=\pm}\frac{ss'}{4}
      \langle
       n_{i:n,s}n_{j:n+r,s'}
      \rangle_{\rm GS}\,,
   \nonumber
\end{eqnarray}
where $\langle\cdots\rangle_{\rm GS}$ denotes the canonical average
at absolute zero point and may approximately be identified with the
Monte Carlo sampling without any global flip along the Trotter
direction in the zero-magnetization subspace at a sufficiently low
temperature \cite{Yama70}.
In Fig. \ref{F:SC}(a), no spin moment is essentially observed on the
X sublattice, whereas there clearly exist antiferromagnetic spin-spin
correlations on the M$_2$ sublattice.
In Fig. \ref{F:SC}(b), we find weak but surviving antiferromagnetic
spin-spin correlations on the X sublattice.
In the X-SDW state, the spin moments induced on the X sites are not
so large as those on the M$_2$ moieties in the M-SDW state and their
one-dimensional stabilization is less pronounced.
However, observing $f^{\rm MXM}(r)$ as well, we may fully be
convinced of the symmetry properties illustrated in Fig.
\ref{F:DW}(g) and the two-band character of X-SDW.

   Finally in this section, let us turn back to Figs.
\ref{F:snapMCDW}-\ref{F:snapXSDW} and take a look at
finite-temperature snapshots.
The spin correlations do not survive thermal fluctuations at all
within one dimension, while the charge correlations are rather tough
against increasing temperature.
Such an aspect is more clearly demonstrated by M-CDW than by X-CDW.
On the other hand, we have learned in Fig. \ref{F:PhD} that X-CDW
much more stably exists in the ground-state phase diagram.
There may be a viewpoint that M-CDW is tough against thermal
fluctuations but is sensitive to electronic correlations, while
X-CDW is tough against quantum fluctuations but less survives thermal
fluctuations.
The M-CDW-type ground state is widely observed for the pop series of
MMX chains \cite{Kurm20,Clar09,Butl55,Kimu40,Wada95}, whereas the
X-CDW-type one is found for the dta complexes \cite{Kita11,Kita68}.
In (NH$_4$)$_4$[Pt$_2$(pop)$_4$X], the valence structure of the M-CDW
type remains stable, in spite of minor formation of paramagnetic
sites, up to room temperature \cite{Kimu40}.
On the other hand, Pt$_2$(dta)$_4$I has been reported to undergoes a
phase transition from X-CDW to BOW around 80 K, far prior to the
metallic behavior above 300 K \cite{Kita68}.
These observations are qualitatively consistent with the present
calculations.
More extensive finite-temperature calculations will be reported
elsewhere.

\section{Summary and Discussion}

\begin{figure}
\begin{flushleft}
\ \mbox{\psfig{figure=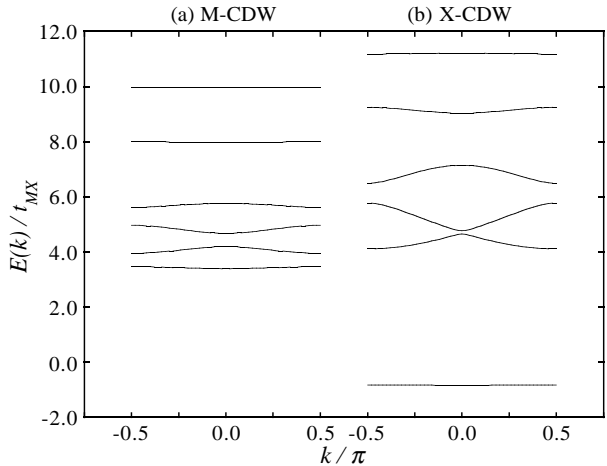,width=80mm,angle=0}}
\end{flushleft}
\vskip 0mm
\caption{Band dispersions peculiar to M-CDW (a) and X-CDW (b), which
         are calculated within the Hartree-Fock approximation under
         the same parametrizations as Figs. 6 and 7, respectively.}
\label{F:BD}
\end{figure}

   We have investigated the electronic properties of MMX chains both
analytically and numerically with particular emphasis on the
$d$-$p$-hybridization effect on competing charge- or spin-ordered
states.
Density-wave states of the M-type, M-CDW and M-SDW, may be described
by a single-band model, whereas $p$-electrons play an essential role
in stabilizing those of the X-type, X-CDW and X-SDW.
Such views of these density-wave states are further supported by
their band dispersions shown in Fig. \ref{F:BD}.
In comparison with M-CDW, X-CDW exhibits much more widespread energy
bands due to an essential hybridization between the $d$ and $p$
orbitals.
Based on the present calculations together with the thus-far reported
experimental observations, we characterize the pop and dta families
of MMX compounds as $d_{z^2}$-single-band and $d$-$p$-hybridized
two-band materials, respectively.
Besides this viewpoint, M-CDW and X-CDW present a few contrasts
between them:
They are, respectively, stable and instable against thermal
fluctuations, while fragile and tough in the ground-state phase
diagram.
\vskip 1mm
\vskip 1mm

   At the naivest consideration, single-band descriptions of MMX
chains may more or less be justified unless ${\mit\Delta}\varepsilon$
is small enough.
However, this is not necessarily the case under the existence of
electronic correlations.
If we employ a single-band Hamiltonian \cite{Kuwa}
%\begin{eqnarray}
%   {\cal H}
%   &=&\frac{K}{2}\sum_{n}
%      \big(l_{1:n}^2+l_{2:n}^2\big)
%    - \beta\sum_{n,s}
%      \big(l_{1:n}n_{1:n,s}-l_{2:n}n_{2:n,s}\big)
%      \nonumber \\
%   &-&\sum_{n,s}
%      \Bigl[t_{\rm MXM}-\alpha\big(l_{1:n}-l_{2:n}\big)\Bigr]
%      \big(a_{1:n,s}^\dagger a_{2:n,s}+{\rm H.c.}\big)
%    - t_{\rm MM}\sum_{n,s}
%      \big(a_{1:n,s}^\dagger a_{2:n-1,s}+{\rm H.c.}\big)
%      \nonumber \\
%   &+&U_{\rm M}\sum_{n}
%      \big(n_{1:n,+}n_{1:n,-}+n_{2:n,+}n_{2:n,-}\big)
%      \nonumber \\
%    + \sum_{n,s,s'}
%      V_{\rm MM}\,n_{1:n,s}n_{2:n-1,s'}\,,
%   \label{E:sH}
%\end{eqnarray}
\begin{eqnarray}
   {\cal H}
   &=&\frac{K}{2}\sum_{n}
      \big(l_{1:n}^2+l_{2:n}^2\big)
    - \beta\sum_{n,s}
      \big(l_{1:n}n_{1:n,s}-l_{2:n}n_{2:n,s}\big)
      \nonumber \\
   &-&\sum_{n,s}
      \Bigl[t_{\rm MXM}-\alpha\big(l_{1:n}-l_{2:n}\big)\Bigr]
      \big(a_{1:n,s}^\dagger a_{2:n,s}+{\rm H.c.}\big)
      \nonumber \\
   &-&t_{\rm MM}\sum_{n,s}
      \big(a_{1:n,s}^\dagger a_{2:n-1,s}+{\rm H.c.}\big)
      \nonumber \\
   &+&U_{\rm M}\sum_{n}
      \big(n_{1:n,+}n_{1:n,-}+n_{2:n,+}n_{2:n,-}\big)
      \nonumber \\
   &+&\sum_{n,s,s'}
      V_{\rm MM}\,n_{1:n,s}n_{2:n-1,s'}\,,
   \label{E:sH}
\end{eqnarray}
the resultant phase diagrams are qualitatively consistent with the
present calculations without any Coulomb interaction.
Once we turn on $U_{\rm X}$ in the two-band Hamiltonian (\ref{E:H}),
the single-band model (\ref{E:sH}), no matter how well all the
parameters are renormalized, considerably underestimates the
stability of the M$_2$-sublattice-dimerized state in comparison with
the present calculations.
This fact is quite suggestive considering that $U_{\rm X}$ plays in
favor of the oxidation of X$^-$ ions.
Though assignment of any parameter is not yet successfully done in
MMX compounds, $U_{\rm M}$ and $U_{\rm X}$ must be larger than the
transfer energy and be of the order of eV on the analogy of MX
compounds \cite{Gamm08}.
Thus an explicit treatment of the X $p_z$ orbitals is indispensable
to the dta complexes.

   Experimental study on halogen-bridged multinuclear metal complexes
is still in its early stage.
X-ray photoelectron spectroscopy measurements on Pt$_2$(dta)$_4$I
\cite{Bell44,Kita68} are now under controversy.
Polymorphism inherent in the pop-family compounds, partly originating
from the contained water molecules, sometimes prevent us from
observing the proper behavior of these materials.
We hope the present research will stimulate further measurements and
lead to close collaboration between theoretical and experimental
investigations.

\acknowledgments

   The author is grateful to A. R. Bishop, K. Yonemitsu, and M.
Kuwabara for useful discussions.
He is thankful to H. Okamoto, H. Kitagawa, and K. Kanoda as well for
helpful comments on MMX materials and measurements on them.
This work was supported by the Japanese Ministry of Education,
Science, and Culture and by the Sanyo-Broadcasting Foundation for
Science and Culture.
The numerical calculation was done using the facility of the
Supercomputer Center, Institute for Solid State Physics, University
of Tokyo.

\begin{appendix}
\section{Irreducible Hamiltonians}
\label{A:h}

\begin{eqnarray}
   h_{{\mit\Gamma}A}^\lambda
    &=&
      a_{{\mit\Gamma}A}^\lambda\sum_{k,s,s'}
      \big(
       a_{1:k,s}^\dagger a_{1:k,s'}
      +a_{2:k,s}^\dagger a_{2:k,s'}
      \big)\sigma_{ss'}^\lambda
    \nonumber \\
    &+&
      b_{{\mit\Gamma}A}^\lambda\sum_{k,s,s'}
       a_{3:k,s}^\dagger a_{3:k,s'}\sigma_{ss'}^\lambda
    \nonumber \\
    &+&
      c_{{\mit\Gamma}A}^\lambda\sum_{k,s,s'}
      \Bigl[
       {\rm e}^{-{\rm i}k/3}
       a_{1:k,s}^\dagger a_{2:k,s'}\sigma_{ss'}^\lambda
      +{\rm H.c.}
      \Bigr]
    \nonumber \\
    &+&
      d_{{\mit\Gamma}A}^\lambda\sum_{k,s,s'}
      \Bigl[
      \big(
       {\rm e}^{ {\rm i}k/3}
       a_{1:k,s}^\dagger a_{3:k,s'}
    \nonumber \\
    &&\quad
      +{\rm e}^{-{\rm i}k/3}
       a_{2:k,s}^\dagger a_{3:k,s'}
      \big)\sigma_{ss'}^\lambda
      +{\rm H.c.}
      \Bigr]\,,
    \label{E:hGA}
    \\
   h_{{\mit\Gamma}B}^\lambda
    &=&
      a_{{\mit\Gamma}B}^\lambda\sum_{k,s,s'}
      \big(
       a_{1:k,s}^\dagger a_{1:k,s'}
      -a_{2:k,s}^\dagger a_{2:k,s'}
      \big)\sigma_{ss'}^\lambda
    \nonumber \\
    &+&
      b_{{\mit\Gamma}B}^\lambda\sum_{k,s,s'}
      \Bigl[
      \big(
       {\rm e}^{ {\rm i}k/3}
       a_{1:k,s}^\dagger a_{3:k,s'}
    \nonumber \\
    &&\quad
      -{\rm e}^{-{\rm i}k/3}
       a_{2:k,s}^\dagger a_{3:k,s'}
      \big)\sigma_{ss'}^\lambda
      +{\rm H.c.}
      \Bigr]\,,
    \label{E:hGB}
    \\
   h_{XA}^\lambda
    &=&
      a_{XA}^\lambda\sum_{k,s,s'}
      \big(
       {\rm e}^{ {\rm i}\pi/3}
       a_{1:k+\pi,s}^\dagger a_{1:k,s'}
    \nonumber \\
    &&\quad
      +{\rm e}^{-{\rm i}\pi/3}
       a_{2:k+\pi,s}^\dagger a_{2:k,s'}
      \big)\sigma_{ss'}^\lambda
    \nonumber \\
    &+&
      b_{XA}^\lambda\sum_{k,s,s'}
       a_{3:k+\pi,s}^\dagger a_{3:k,s'}\sigma_{ss'}^\lambda
    \nonumber \\
    &+&
      c_{XA}^\lambda\sum_{k,s,s'}
      \Bigl[
      \big(
       {\rm e}^{ {\rm i}(k+\pi)/3}
       a_{1:k+\pi,s}^\dagger a_{3:k,s'}
    \nonumber \\
    &&\quad
      +{\rm e}^{-{\rm i}(k+\pi)/3}
       a_{2:k+\pi,s}^\dagger a_{3:k,s'}
      \big)\sigma_{ss'}^\lambda
      +{\rm H.c.}
      \Bigr]\,,
    \label{E:hXA}
    \\
   h_{XB}^\lambda
    &=&
      a_{XB}^\lambda\sum_{k,s,s'}
      \big(
       {\rm e}^{ {\rm i}\pi/3}
       a_{1:k+\pi,s}^\dagger a_{1:k,s'}
    \nonumber \\
    &&\quad
      -{\rm e}^{-{\rm i}\pi/3}
       a_{2:k+\pi,s}^\dagger a_{2:k,s'}
      \big)\sigma_{ss'}^\lambda
    \nonumber \\
    &+&
      b_{XB}^\lambda\sum_{k,s,s'}
      \Bigl[
       {\rm e}^{-{\rm i}(k-\pi)/3}
       a_{1:k+\pi,s}^\dagger a_{2:k,s'}\sigma_{ss'}^\lambda
      +{\rm H.c.}
      \Bigr]
    \nonumber \\
    &+&
      c_{XB}^\lambda\sum_{k,s,s'}
      \Bigl[
      \big(
       {\rm e}^{ {\rm i}(k+\pi)/3}
       a_{1:k+\pi,s}^\dagger a_{3:k,s'}
    \nonumber \\
    &&\quad
      -{\rm e}^{-{\rm i}(k+\pi)/3}
       a_{2:k+\pi,s}^\dagger a_{3:k,s'}
      \big)\sigma_{ss'}^\lambda
      +{\rm H.c.}
      \Bigr]\,,
    \label{E:hXB}
\end{eqnarray}
where
\begin{eqnarray}
  &&a_{{\mit\Gamma}A}^0
    =\varepsilon_{\rm M}-\frac{U_{\rm M}}{2}
    +\Bigl(
      \frac{U_{\rm M}}{2N}
     +\frac{2V_{\rm MM}}{N}
     \Bigr)
     \sum_k
     \Bigl[
      \rho_{11}^0({\mit\Gamma};k)
    \nonumber \\
  &&\qquad
     +\rho_{22}^0({\mit\Gamma};k)
     \Bigr]
    +\frac{4V_{\rm MX}}{N}
     \sum_k
     \rho_{33}^0({\mit\Gamma};k)\,,
    \nonumber \\
  &&b_{{\mit\Gamma}A}^0
    =\varepsilon_{\rm X}-\frac{U_{\rm X}}{2}
    +\frac{4V_{\rm MX}}{N}
     \sum_k
     \Bigl[
      \rho_{11}^0({\mit\Gamma};k)
    \nonumber \\
  &&\qquad
     +\rho_{22}^0({\mit\Gamma};k)
     \Bigr]
    +\frac{U_{\rm X}}{N}
     \sum_k
     \rho_{33}^0({\mit\Gamma};k)\,,
    \nonumber \\
  &&c_{{\mit\Gamma}A}^0
    =-t_{\rm MM}-\frac{2V_{\rm MM}}{N}
     \sum_k
     {\rm e}^{{\rm i}k/3}\rho_{12}^0({\mit\Gamma};k)\,,
    \nonumber \\
  &&d_{{\mit\Gamma}A}^0
    =-t_{\rm MX}-\frac{V_{\rm MX}}{N}
     \sum_k
     \Bigl[
      {\rm e}^{-{\rm i}k/3}\rho_{13}^0({\mit\Gamma};k)
    \nonumber \\
  &&\qquad
     +{\rm e}^{ {\rm i}k/3}\rho_{23}^0({\mit\Gamma};k)
     \Bigr]\,,
    \label{E:SCFGA0}
    \\
  &&a_{{\mit\Gamma}B}^0
    =\Bigl(
      \frac{U_{\rm M}}{2N}
     -\frac{2V_{\rm MM}}{N}
     -\frac{\beta^2}{KN}
     \Bigr)
     \sum_k
     \Bigl[
      \rho_{11}^0({\mit\Gamma};k)
    \nonumber \\
  &&\qquad
     -\rho_{22}^0({\mit\Gamma};k)
     \Bigr]
    +\frac{2\alpha\beta}{KN}
     \sum_k
     \Bigl[
      {\rm e}^{-{\rm i}k/3}
      \rho_{13}^0({\mit\Gamma};k)
    \nonumber \\
  &&\qquad
     -{\rm e}^{ {\rm i}k/3}
      \rho_{23}^0({\mit\Gamma};k)
     \Bigr]\,,
    \nonumber \\
  &&b_{{\mit\Gamma}B}^0
    =-
     \Bigl(
      \frac{V_{\rm MX}}{N}
     -\frac{2\alpha^2}{KN}
     \Bigr)
     \sum_k
     \Bigl[
      {\rm e}^{-{\rm i}k/3}
      \rho_{13}^0({\mit\Gamma};k)
    \nonumber \\
  &&\qquad
     -{\rm e}^{ {\rm i}k/3}
      \rho_{23}^0({\mit\Gamma};k)
     \Bigr]
    +\frac{\alpha\beta}{KN}
     \sum_k
     \Bigl[
      \rho_{11}^0({\mit\Gamma};k)
    \nonumber \\
  &&\qquad
     -\rho_{22}^0({\mit\Gamma};k)
     \Bigr]\,,
    \label{E:SCFGB0}
    \\
  &&a_{XA}^0
    =\Bigl(
      \frac{U_{\rm M}}{2N}
     +\frac{2V_{\rm MM}}{N}
     -\frac{\beta^2}{KN}
     \Bigr)
     \sum_k
     \Bigl[
      {\rm e}^{ {\rm i}\pi/3}
      \rho_{11}^0(X;k)
    \nonumber \\
  &&\qquad
     +{\rm e}^{-{\rm i}\pi/3}
      \rho_{22}^0(X;k)
     \Bigr]
    +\frac{4V_{\rm MX}}{N}
     \sum_k
     \rho_{33}^0(X,k)
    \nonumber \\
  &&\qquad
    +\frac{2\alpha\beta}{KN}
     \sum_k
     \Bigl[
      {\rm e}^{-{\rm i}k/3}
      \rho_{13}^0(X;k)
     +{\rm e}^{ {\rm i}k/3}
      \rho_{23}^0(X;k)
     \Bigr]\,,
    \nonumber \\
  &&b_{XA}^0
    =\frac{4V_{\rm MX}}{N}
     \sum_k
     \Bigl[
      {\rm e}^{ {\rm i}\pi/3}
      \rho_{11}^0(X;k)
     +{\rm e}^{-{\rm i}\pi/3}
      \rho_{22}^0(X;k)
     \Bigr]
    \nonumber \\
  &&\qquad
    +\frac{U_{\rm X}}{N}
     \sum_k
     \rho_{33}^0(X;k)\,,
    \nonumber \\
  &&c_{XA}^0
    =
    -\Bigl(
      \frac{V_{\rm MX}}{N}
     +\frac{2\alpha^2}{KN}
     \Bigr)
     \sum_k
     \Bigl[
      {\rm e}^{-{\rm i}k/3}
      \rho_{13}^0(X;k)
    \nonumber \\
  &&\qquad
     +{\rm e}^{ {\rm i}k/3}
      \rho_{23}^0(X;k)
     \Bigr]
    +\frac{\alpha\beta}{KN}
     \sum_k
     \Bigl[
      {\rm e}^{ {\rm i}\pi/3}
      \rho_{11}^0(X;k)
    \nonumber \\
  &&\qquad
     +{\rm e}^{-{\rm i}\pi/3}
      \rho_{22}^0(X;k)
     \Bigr]\,,
    \label{E:SCFXA0}
    \\
  &&a_{XB}^0
    =\Bigl(
      \frac{U_{\rm M}}{2N}
     -\frac{2V_{\rm MM}}{N}
     -\frac{\beta^2}{KN}
     \Bigr)
     \sum_k
     \Bigl[
      {\rm e}^{ {\rm i}\pi/3}
      \rho_{11}^0(X;k)
    \nonumber \\
  &&\qquad
     -{\rm e}^{-{\rm i}\pi/3}
      \rho_{22}^0(X;k)
     \Bigr]
    +\frac{2\alpha\beta}{KN}
     \sum_k
     \Bigl[
      {\rm e}^{-{\rm i}k/3}
      \rho_{13}^0(X;k)
    \nonumber \\
  &&\qquad
     -{\rm e}^{ {\rm i}k/3}
      \rho_{23}^0(X;k)
     \Bigr]\,,
    \nonumber \\
  &&b_{XB}^0
    =\frac{2V_{\rm MM}}{N}
     \sum_k
     {\rm e}^{{\rm i}(k-\pi)/3}
     \rho_{12}^0(X;B)\,,
    \nonumber \\
  &&c_{XB}^0
    =
    -\Bigl(
      \frac{V_{\rm MX}}{N}
     +\frac{2\alpha^2}{KN}
     \Bigr)
     \sum_k
     \Bigl[
      {\rm e}^{-{\rm i}k/3}
      \rho_{13}^0(X;k)
    \nonumber \\
  &&\qquad
     -{\rm e}^{ {\rm i}k/3}
      \rho_{23}^0(X;k)
     \Bigr]
    +\frac{\alpha\beta}{KN}
     \sum_k
     \Bigl[
      {\rm e}^{ {\rm i}\pi/3}
      \rho_{11}^0(X;k)
    \nonumber \\
  &&\qquad
     -{\rm e}^{-{\rm i}\pi/3}
      \rho_{22}^0(X;k)
     \Bigr]\,,
    \label{E:SCFXB0}
    \\
  &&a_{{\mit\Gamma}A}^z
    =
    -\frac{U_{\rm M}}{2N}
     \sum_k
     \Bigl[
      \rho_{11}^z({\mit\Gamma};k)
     +\rho_{22}^z({\mit\Gamma};k)
     \Bigr]\,,
    \nonumber \\
  &&b_{{\mit\Gamma}A}^z
    =
    -\frac{U_{\rm X}}{N}
     \sum_k
     \rho_{33}^z({\mit\Gamma};k)\,,
    \nonumber \\
  &&c_{{\mit\Gamma}A}^z
    =
    -\frac{2V_{\rm MM}}{N}
     \sum_k
     {\rm e}^{{\rm i}k/3}
     \rho_{12}^z({\mit\Gamma};k)\,,
    \nonumber \\
  &&d_{{\mit\Gamma}A}^z
    =
    -\frac{V_{\rm MX}}{N}
     \sum_k
     \Bigl[
      {\rm e}^{-{\rm i}k/3}
      \rho_{13}^z({\mit\Gamma};k)
     +{\rm e}^{ {\rm i}k/3}
      \rho_{23}^z({\mit\Gamma};k)
     \Bigr]\,,
    \nonumber \\
  &&\qquad
    \label{E:SCFGAz}
    \\
  &&a_{{\mit\Gamma}B}^z
    =
    -\frac{U_{\rm M}}{2N}
     \sum_k
     \Bigl[
      \rho_{11}^z({\mit\Gamma};k)
     -\rho_{22}^z({\mit\Gamma};k)
     \Bigr]\,,
    \nonumber \\
  &&b_{{\mit\Gamma}B}^z
    =
    -\frac{V_{\rm MX}}{N}
     \sum_k
     \Bigl[
      {\rm e}^{-{\rm i}k/3}
      \rho_{13}^z({\mit\Gamma};k)
     -{\rm e}^{ {\rm i}k/3}
      \rho_{23}^z({\mit\Gamma};k)
     \Bigr]\,,
    \nonumber \\
  &&\qquad
    \label{E:SCFGBz}
    \\
  &&a_{XA}^z
    =
    -\frac{U_{\rm M}}{2N}
     \sum_k
     \Bigl[
      {\rm e}^{ {\rm i}\pi/3}
      \rho_{11}^z(X;k)
     +{\rm e}^{-{\rm i}\pi/3}
      \rho_{22}^z(X;k)
     \Bigr]\,,
    \nonumber \\
  &&b_{XA}^z
    =
    -\frac{U_{\rm X}}{N}
     \sum_k
     \rho_{33}^z(X;k)\,,
    \nonumber \\
  &&c_{XA}^z
    =
    -\frac{V_{\rm MX}}{N}
     \sum_k
     \Bigl[
      {\rm e}^{-{\rm i}k/3}
      \rho_{13}^z(X;k)
     +{\rm e}^{ {\rm i}k/3}
      \rho_{23}^z(X;k)
     \Bigr]\,,
    \nonumber \\
  &&\qquad
    \label{E:SCFXAz}
    \\
  &&a_{XB}^z
    =
    -\frac{U_{\rm M}}{2N}
     \sum_k
     \Bigl[
      {\rm e}^{ {\rm i}\pi/3}
      \rho_{11}^z(X;k)
     -{\rm e}^{-{\rm i}\pi/3}
      \rho_{22}^z(X;k)
     \Bigr]\,,
    \nonumber \\
  &&b_{XB}^z
    =\frac{V_{\rm MM}}{N}
     \sum_k
     {\rm e}^{{\rm i}(k-\pi)/3}
     \rho_{12}^z(X;k)\,,
    \nonumber \\
  &&c_{XB}^z
    =
    -\frac{V_{\rm MX}}{N}
     \sum_k
     \Bigl[
      {\rm e}^{-{\rm i}k/3}
      \rho_{13}^z(X;k)
     -{\rm e}^{ {\rm i}k/3}
      \rho_{23}^z(X;k)
     \Bigr]\,.
    \nonumber \\
  &&\qquad
    \label{E:SCFXBz}
\end{eqnarray}
In obtaining the spatial-symmetry-definite order parameters
(\ref{E:SCFGA0})-(\ref{E:SCFXBz}), the lattice distortion has been
described in terms of the electron density matrices so as to minimize
the Hartree-Fock energy
\begin{eqnarray}
  &&
  \langle{\cal H}\rangle_{\rm HF}
   =2\sum_{i,j}\sum_k
    \langle i:k|t|j:k \rangle
    \rho_{ji}^0({\mit\Gamma};k)
  \nonumber \\
  &&\quad
   +2\sum_{i,j}\sum_k
    \langle i:k+\pi|t|j:k \rangle
    \rho_{ji}^0(X;k)
  \nonumber \\
  &&\quad
   +\sum_{i,j,m,n}\sum_{k,k'}
    \big(
     2\langle i:k;m:k'|v|j:k;n:k'\rangle
  \nonumber \\
  &&\qquad
     -\langle i:k;m:k'|v|n:k';j:k\rangle
    \big)
    \rho_{ji}^0({\mit\Gamma};k)
    \rho_{nm}^0({\mit\Gamma};k')
  \nonumber \\
  &&\quad
   +\sum_{i,j,m,n}\sum_{k,k'}
    \big(
     2\langle i:k+\pi;m:k'|v|j:k;n:k'+\pi\rangle
  \nonumber \\
  &&\qquad
     -\langle i:k+\pi;m:k'|v|n:k'+\pi;j:k\rangle
    \big)
  \nonumber \\
  &&\qquad\times
    {\rm e}^{ (2\pi/3)i\delta_{m1}}
    {\rm e}^{-(2\pi/3)i\delta_{m2}}
    \rho_{ji}^0({\mit\Gamma};k)
    \rho_{nm}^0({\mit\Gamma};k'+\pi)
  \nonumber \\
  &&\quad
   -\sum_{i,j,m,n}\sum_{k,k'}
    \langle i:k;m:k'|v|n:k';j:k\rangle
  \nonumber \\
  &&\qquad\times
    \rho_{ji}^z({\mit\Gamma},k)
    \rho_{nm}^z({\mit\Gamma},k')
  \nonumber \\
  &&\quad
   -\sum_{i,j,m,n}\sum_{k,k'}
    \langle i:k+\pi;m:k'|v|n:k'+\pi;j:k\rangle
  \nonumber \\
  &&\qquad\times
    {\rm e}^{ (2\pi/3)i\delta_{m1}}
    {\rm e}^{-(2\pi/3)i\delta_{m2}}
    \rho_{ji}^z({\mit\Gamma},k)
    \rho_{nm}^z({\mit\Gamma},k'+\pi)
  \nonumber \\
  &&\quad
   +\frac{K}{2}\sum_{q=0,\pi}
    \Bigl[
     \big(
      u_{3:q}-{\rm e}^{ {\rm i}q/3}u_{1:q}
     \big)^2
    +\big(
      u_{3:q}-{\rm e}^{-{\rm i}q/3}u_{2:q}
     \big)^2
    \Bigr]\,.
  \nonumber \\
  &&\quad
  \label{E:EHF}
\end{eqnarray}
%\begin{eqnarray}
%  &&
%  \langle{\cal H}\rangle_{\rm HF}
%   =2\sum_{i,j}\sum_k
%    \langle i:k|t|j:k \rangle
%    \rho_{ji}^0({\mit\Gamma};k)
%   +2\sum_{i,j}\sum_k
%    \langle i:k+\pi|t|j:k \rangle
%    \rho_{ji}^0(X;k)
%  \nonumber \\
%  &&\quad
%   +\sum_{i,j,m,n}\sum_{k,k'}
%    \big(
%     2\langle i:k;m:k'|v|j:k;n:k'\rangle
%     -\langle i:k;m:k'|v|n:k';j:k\rangle
%    \big)
%    \rho_{ji}^0({\mit\Gamma};k)
%    \rho_{nm}^0({\mit\Gamma};k')
%  \nonumber \\
%  &&\quad
%   +\sum_{i,j,m,n}\sum_{k,k'}
%    \big(
%     2\langle i:k+\pi;m:k'|v|j:k;n:k'+\pi\rangle
%     -\langle i:k+\pi;m:k'|v|n:k'+\pi;j:k\rangle
%    \big)
%  \nonumber \\
%  &&\qquad\times
%    {\rm e}^{ (2\pi/3)i\delta_{m1}}
%    {\rm e}^{-(2\pi/3)i\delta_{m2}}
%    \rho_{ji}^0({\mit\Gamma};k)
%    \rho_{nm}^0({\mit\Gamma};k'+\pi)
%  \nonumber \\
%  &&\quad
%   -\sum_{i,j,m,n}\sum_{k,k'}
%    \langle i:k;m:k'|v|n:k';j:k\rangle
%    \rho_{ji}^z({\mit\Gamma},k)
%    \rho_{nm}^z({\mit\Gamma},k')
%  \nonumber \\
%  &&\quad
%   -\sum_{i,j,m,n}\sum_{k,k'}
%    \langle i:k+\pi;m:k'|v|n:k'+\pi;j:k\rangle
%    {\rm e}^{ (2\pi/3)i\delta_{m1}}
%    {\rm e}^{-(2\pi/3)i\delta_{m2}}
%    \rho_{ji}^z({\mit\Gamma},k)
%    \rho_{nm}^z({\mit\Gamma},k'+\pi)
%  \nonumber \\
%  &&\quad
%   +\frac{K}{2}\sum_{q=0,\pi}
%    \Bigl[
%     \big(
%      u_{3:q}-{\rm e}^{ {\rm i}q/3}u_{1:q}
%     \big)^2
%    +\big(
%      u_{3:q}-{\rm e}^{-{\rm i}q/3}u_{2:q}
%     \big)^2
%    \Bigr]\,.
%  \label{E:EHF}
%\end{eqnarray}
Considering that the density matrices should possess the same
symmetry properties as the irreducible Hamiltonian to which they
belong, the coefficients (\ref{E:SCFGA0})-(\ref{E:SCFXBz}) all turn
out to be real.

\section{Characterization of the Density-Wave Solutions}
\label{A:OP}

\noindent
(a) ${\mit\Gamma}A\otimes\check{S}^0\otimes\check{T}^0$
\begin{eqnarray}
  &&
  d_{1:n}=d_{2:n}
   =\frac{2}{N}
    \sum_k
    {\bar\rho}_{11}^0({\mit\Gamma};k)\,,
  \nonumber \\
  &&
  d_{3:n}
   =\frac{2}{N}
    \sum_k
    {\bar\rho}_{33}^0({\mit\Gamma};k)\,,
  \nonumber \\
  &&
  p_{1:n;2:n-1}
   =\frac{2}{N}\sum_k
    \Bigl[
     {\bar  \rho}_{12}^0({\mit\Gamma};k){\rm cos}\frac{k}{3}
    -{\tilde\rho}_{12}^0({\mit\Gamma};k){\rm sin}\frac{k}{3}
    \Bigr]\,,
  \nonumber \\
  &&
  p_{1:n;3:n}=p_{2:n;3:n}
  \nonumber \\
  &&\quad
   =\frac{2}{N}
    \sum_k
    \Bigl[
     {\bar  \rho}_{13}^0({\mit\Gamma};k){\rm cos}\frac{k}{3}
    +{\tilde\rho}_{13}^0({\mit\Gamma};k){\rm sin}\frac{k}{3}
    \Bigr]\,.
  \nonumber
\end{eqnarray}
%\begin{eqnarray}
%  &&
%  d_{1:n}=d_{2:n}
%   =\frac{2}{N}
%    \sum_k
%    {\bar\rho}_{11}^0({\mit\Gamma};k)\,,\ \ 
%  d_{3:n}
%   =\frac{2}{N}
%    \sum_k
%    {\bar\rho}_{33}^0({\mit\Gamma};k)\,,
%  \nonumber \\
%  &&
%  p_{1:n;2:n-1}
%   =\frac{2}{N}\sum_k
%    \Bigl[
%     {\bar  \rho}_{12}^0({\mit\Gamma};k){\rm cos}\frac{k}{3}
%    -{\tilde\rho}_{12}^0({\mit\Gamma};k){\rm sin}\frac{k}{3}
%    \Bigr]\,,
%  \nonumber \\
%  &&
%  p_{1:n;3:n}=p_{2:n;3:n}
%   =\frac{2}{N}
%    \sum_k
%    \Bigl[
%     {\bar  \rho}_{13}^0({\mit\Gamma};k){\rm cos}\frac{k}{3}
%    +{\tilde\rho}_{13}^0({\mit\Gamma};k){\rm sin}\frac{k}{3}
%    \Bigr]\,.
%  \nonumber
%\end{eqnarray}

\noindent
(b) ${\mit\Gamma}B\otimes\check{S}^0\otimes\check{T}^0$
\begin{eqnarray}
  &&
  d_{i:n}
   =\frac{2}{N}
    \sum_k
    {\bar\rho}_{ii}^0({\mit\Gamma};k)\ \ (i=1,2,3)\,,
  \nonumber \\
  &&
  p_{1:n;2:n-1}
   =\frac{2}{N}
    \sum_k
    \Bigl[
     {\bar  \rho}_{12}^0({\mit\Gamma};k){\rm cos}\frac{k}{3}
    -{\tilde\rho}_{12}^0({\mit\Gamma};k){\rm sin}\frac{k}{3}
    \Bigr]\,,
  \nonumber \\
  &&
  p_{1:n;3:n}
   =\frac{2}{N}
    \sum_k
    \Bigl[
     {\bar  \rho}_{13}^0({\mit\Gamma};k){\rm cos}\frac{k}{3}
    +{\tilde\rho}_{13}^0({\mit\Gamma};k){\rm sin}\frac{k}{3}
    \Bigr]\,,
  \nonumber \\
  &&
  p_{2:n;3:n}
   =\frac{2}{N}
    \sum_k
    \Bigl[
     {\bar  \rho}_{23}^0({\mit\Gamma};k){\rm cos}\frac{k}{3}
    -{\tilde\rho}_{23}^0({\mit\Gamma};k){\rm sin}\frac{k}{3}
    \Bigr]\,,
  \nonumber \\
  &&
  u_{3:n}
   =\frac{\beta}{KN}\sum_k
    \Bigl[
     {\bar\rho}_{11}^0({\mit\Gamma};k)
    -{\bar\rho}_{22}^0({\mit\Gamma};k)
    \Bigr]
  \nonumber \\
  &&\quad
   -\frac{2\alpha}{KN}\sum_k
    \Bigl[
     {\bar  \rho}_{13}^0({\mit\Gamma};k){\rm cos}\frac{k}{3}
    +{\tilde\rho}_{13}^0({\mit\Gamma};k){\rm sin}\frac{k}{3}
  \nonumber \\
  &&\quad
    -{\bar  \rho}_{23}^0({\mit\Gamma};k){\rm cos}\frac{k}{3}
    +{\tilde\rho}_{23}^0({\mit\Gamma};k){\rm sin}\frac{k}{3}
    \Bigr]\,.
  \nonumber
\end{eqnarray}
%\begin{eqnarray}
%  &&
%  d_{i:n}
%   =\frac{2}{N}
%    \sum_k
%    {\bar\rho}_{ii}^0({\mit\Gamma};k)\ \ (i=1,2,3)\,,
%  \nonumber \\
%  &&
%  p_{1:n;2:n-1}
%   =\frac{2}{N}
%    \sum_k
%    \Bigl[
%     {\bar  \rho}_{12}^0({\mit\Gamma};k){\rm cos}\frac{k}{3}
%    -{\tilde\rho}_{12}^0({\mit\Gamma};k){\rm sin}\frac{k}{3}
%    \Bigr]\,,
%  \nonumber \\
%  &&
%  p_{1:n;3:n}
%   =\frac{2}{N}
%    \sum_k
%    \Bigl[
%     {\bar  \rho}_{13}^0({\mit\Gamma};k){\rm cos}\frac{k}{3}
%    +{\tilde\rho}_{13}^0({\mit\Gamma};k){\rm sin}\frac{k}{3}
%    \Bigr]\,,
%  \nonumber \\
%  &&
%  p_{2:n;3:n}
%   =\frac{2}{N}
%    \sum_k
%    \Bigl[
%     {\bar  \rho}_{23}^0({\mit\Gamma};k){\rm cos}\frac{k}{3}
%    -{\tilde\rho}_{23}^0({\mit\Gamma};k){\rm sin}\frac{k}{3}
%    \Bigr]\,,
%  \nonumber \\
%  &&
%  u_{3:n}
%   =\frac{\beta}{KN}\sum_k
%    \Bigl[
%     {\bar\rho}_{11}^0({\mit\Gamma};k)
%    -{\bar\rho}_{22}^0({\mit\Gamma};k)
%    \Bigr]
%  \nonumber \\
%  &&\quad
%   -\frac{2\alpha}{KN}\sum_k
%    \Bigl[
%     {\bar  \rho}_{13}^0({\mit\Gamma};k){\rm cos}\frac{k}{3}
%    +{\tilde\rho}_{13}^0({\mit\Gamma};k){\rm sin}\frac{k}{3}
%    -{\bar  \rho}_{23}^0({\mit\Gamma};k){\rm cos}\frac{k}{3}
%    +{\tilde\rho}_{23}^0({\mit\Gamma};k){\rm sin}\frac{k}{3}
%    \Bigr]\,.
%  \nonumber
%\end{eqnarray}

\noindent
(c) $XA           \otimes\check{S}^0\otimes\check{T}^0$
\begin{eqnarray}
  &&
  d_{1:n}=d_{2:n}
   =\frac{2}{N}
    \sum_k
    {\bar\rho}_{11}^0({\mit\Gamma};k)
  \nonumber \\
  &&\quad
   +\frac{2(-1)^n}{N}
    \sum_k
    \Bigl[
     {\bar  \rho}_{11}^0(X;k){\rm cos}\frac{\pi}{3}
    -{\tilde\rho}_{11}^0(X;k){\rm sin}\frac{\pi}{3}
    \Bigr]\,,
  \nonumber \\
  &&
  d_{3:n}
   =\frac{2}{N}
    \sum_k
    {\bar\rho}_{33}^0({\mit\Gamma};k)
   +\frac{2(-1)^n}{N}
    \sum_k
    {\bar\rho}_{33}^0(X;k)\,,
  \nonumber \\
  &&
  p_{1:n;2:n-1}
   =\frac{2}{N}
    \sum_k
    \Bigl[
     {\bar  \rho}_{12}^0({\mit\Gamma};k){\rm cos}\frac{k}{3}
    -{\tilde\rho}_{12}^0({\mit\Gamma};k){\rm sin}\frac{k}{3}
    \Bigr]\,,
  \nonumber \\
  &&
  p_{1:n;3:n}=p_{2:n;3:n}
  \nonumber \\
  &&\quad
   =\frac{2}{N}
    \sum_k
     \Bigl[
      {\bar  \rho}_{13}^0({\mit\Gamma};k){\rm cos}\frac{k}{3}
     +{\tilde\rho}_{13}^0({\mit\Gamma};k){\rm sin}\frac{k}{3}
     \Bigr]
  \nonumber \\
  &&\quad
   +\frac{2(-1)^n}{N}
    \sum_k
     \Bigl[
      {\bar  \rho}_{13}^0(X;k){\rm cos}\frac{k}{3}
     +{\tilde\rho}_{13}^0(X;k){\rm sin}\frac{k}{3}
     \Bigr]\,,
  \nonumber \\
  &&
  u_{1:n}=-u_{2:n}
   =(-1)^n
  \nonumber \\
  &&\quad\times
    \Big\{
     \frac{4\alpha}{KN}
     \sum_k
     \Bigl[
      {\bar  \rho}_{13}^0(X;k){\rm cos}\frac{k}{3}
     +{\tilde\rho}_{13}^0(X;k){\rm sin}\frac{k}{3}
     \Bigr]
  \nonumber \\
  &&\quad
    -\frac{2\beta}{KN}
     \sum_k
     \Bigl[
      {\bar  \rho}_{13}^0(X;k){\rm cos}\frac{\pi}{3}
     -{\tilde\rho}_{13}^0(X;k){\rm sin}\frac{\pi}{3}
     \Bigr]
    \Bigr\}\,.
  \nonumber
\end{eqnarray}
%\begin{eqnarray}
%  &&
%  d_{1:n}=d_{2:n}
%   =\frac{2}{N}
%    \sum_k
%    {\bar\rho}_{11}^0({\mit\Gamma};k)
%   +\frac{2(-1)^n}{N}
%    \sum_k
%    \Bigl[
%     {\bar  \rho}_{11}^0(X;k){\rm cos}\frac{\pi}{3}
%    -{\tilde\rho}_{11}^0(X;k){\rm sin}\frac{\pi}{3}
%    \Bigr]\,,
%  \nonumber \\
%  &&
%  d_{3:n}
%   =\frac{2}{N}
%    \sum_k
%    {\bar\rho}_{33}^0({\mit\Gamma};k)
%   +\frac{2(-1)^n}{N}
%    \sum_k
%    {\bar\rho}_{33}^0(X;k)\,,
%  \nonumber \\
%  &&
%  p_{1:n;2:n-1}
%   =\frac{2}{N}
%    \sum_k
%    \Bigl[
%     {\bar  \rho}_{12}^0({\mit\Gamma};k){\rm cos}\frac{k}{3}
%    -{\tilde\rho}_{12}^0({\mit\Gamma};k){\rm sin}\frac{k}{3}
%    \Bigr]\,,
%  \nonumber \\
%  &&
%  p_{1:n;3:n}=p_{2:n;3:n}
%   =\frac{2}{N}
%    \sum_k
%     \Bigl[
%      {\bar  \rho}_{13}^0({\mit\Gamma};k){\rm cos}\frac{k}{3}
%     +{\tilde\rho}_{13}^0({\mit\Gamma};k){\rm sin}\frac{k}{3}
%     \Bigr]
%  \nonumber \\
%  &&\quad
%   +\frac{2(-1)^n}{N}
%    \sum_k
%     \Bigl[
%      {\bar  \rho}_{13}^0(X;k){\rm cos}\frac{k}{3}
%     +{\tilde\rho}_{13}^0(X;k){\rm sin}\frac{k}{3}
%     \Bigr]\,,
%  \nonumber \\
%  &&
%  u_{1:n}=-u_{2:n}
%   =(-1)^n
%    \Big\{
%     \frac{4\alpha}{KN}
%     \sum_k
%     \Bigl[
%      {\bar  \rho}_{13}^0(X;k){\rm cos}\frac{k}{3}
%     +{\tilde\rho}_{13}^0(X;k){\rm sin}\frac{k}{3}
%     \Bigr]
%  \nonumber \\
%  &&\quad
%    -\frac{2\beta}{KN}
%     \sum_k
%     \Bigl[
%      {\bar  \rho}_{13}^0(X;k){\rm cos}\frac{\pi}{3}
%     -{\tilde\rho}_{13}^0(X;k){\rm sin}\frac{\pi}{3}
%     \Bigr]
%    \Bigr\}\,.
%  \nonumber
%\end{eqnarray}

\noindent
(d) $XB           \otimes\check{S}^0\otimes\check{T}^0$
\begin{eqnarray}
  &&
  d_{1:n}
   =\frac{2}{N}
    \sum_k
    {\bar\rho}_{11}^0({\mit\Gamma};k)
  \nonumber \\
  &&\quad
   +\frac{2(-1)^n}{N}
    \sum_k
    \Bigl[
     {\bar  \rho}_{11}^0(X;k){\rm cos}\frac{\pi}{3}
    -{\tilde\rho}_{11}^0(X;k){\rm sin}\frac{\pi}{3}
    \Bigr]\,,
  \nonumber \\
  &&
  d_{2:n}
   =\frac{2}{N}
    \sum_k
    {\bar\rho}_{11}^0({\mit\Gamma};k)
  \nonumber \\
  &&\quad
   -\frac{2(-1)^n}{N}
    \sum_k
    \Bigl[
     {\bar  \rho}_{11}^0(X;k){\rm cos}\frac{\pi}{3}
    -{\tilde\rho}_{11}^0(X;k){\rm sin}\frac{\pi}{3}
    \Bigr]\,,
  \nonumber \\
  &&
  d_{3:n}
   =\frac{2}{N}
    \sum_k
    {\bar\rho}_{33}^0({\mit\Gamma};k)\,,
  \nonumber \\
  &&
  p_{1:n;2:n-1}
   =\frac{2}{N}
    \sum_k
    \Bigl[
     {\bar  \rho}_{12}^0({\mit\Gamma};k){\rm cos}\frac{k}{3}
    -{\tilde\rho}_{12}^0({\mit\Gamma};k){\rm sin}\frac{k}{3}
    \Bigr]
  \nonumber \\
  &&\quad
   -\frac{2(-1)^n}{N}
    \sum_k
  \nonumber \\
  &&\quad\times
    \Bigl[
     {\bar  \rho}_{12}^0(X;k){\rm cos}\frac{k-\pi}{3}
    -{\tilde\rho}_{12}^0(X;k){\rm sin}\frac{k-\pi}{3}
    \Bigr]\,,
  \nonumber \\
  &&
  p_{1:n;3:n}
   =\frac{2}{N}
    \sum_k
     \Bigl[
      {\bar  \rho}_{13}^0({\mit\Gamma};k){\rm cos}\frac{k}{3}
     +{\tilde\rho}_{13}^0({\mit\Gamma};k){\rm sin}\frac{k}{3}
     \Bigr]
  \nonumber \\
  &&\quad
   +\frac{2(-1)^n}{N}
    \sum_k
     \Bigl[
      {\bar  \rho}_{13}^0(X;k){\rm cos}\frac{k}{3}
     +{\tilde\rho}_{13}^0(X;k){\rm sin}\frac{k}{3}
     \Bigr]\,,
  \nonumber \\
  &&
  p_{2:n;3:n}
   =\frac{2}{N}
    \sum_k
     \Bigl[
      {\bar  \rho}_{13}^0({\mit\Gamma};k){\rm cos}\frac{k}{3}
     +{\tilde\rho}_{13}^0({\mit\Gamma};k){\rm sin}\frac{k}{3}
     \Bigr]
  \nonumber \\
  &&\quad
   -\frac{2(-1)^n}{N}
    \sum_k
     \Bigl[
      {\bar  \rho}_{13}^0(X;k){\rm cos}\frac{k}{3}
     +{\tilde\rho}_{13}^0(X;k){\rm sin}\frac{k}{3}
     \Bigr]\,,
  \nonumber \\
  &&
  u_{3:n}
   =(-1)^n
  \nonumber \\
  &&\quad\times
    \Big\{
     \frac{4\alpha}{KN}
     \sum_k
     \Bigl[
      {\bar  \rho}_{13}^0(X;k){\rm cos}\frac{k}{3}
     +{\tilde\rho}_{13}^0(X;k){\rm sin}\frac{k}{3}
     \Bigr]
  \nonumber \\
  &&\quad
    -\frac{2\beta}{KN}
     \sum_k
     \Bigl[
      {\bar  \rho}_{13}^0(X;k){\rm cos}\frac{\pi}{3}
     -{\tilde\rho}_{13}^0(X;k){\rm sin}\frac{\pi}{3}
     \Bigr]
    \Bigr\}\,.
  \nonumber
\end{eqnarray}

\noindent
(e) ${\mit\Gamma}A\otimes\check{S}^1\otimes\check{T}^1$
\begin{eqnarray}
  &&
  s_{1:n}^z=s_{2:n}^z
   =\frac{1}{N}
    \sum_k
    {\bar\rho}_{11}^z({\mit\Gamma};k)\,,
  \nonumber \\
  &&
  s_{3:n}^z
   =\frac{1}{N}
    \sum_k
    {\bar\rho}_{33}^z({\mit\Gamma};k)\,,
  \nonumber \\
  &&
  t_{1:n;2:n-1}^z
   =\frac{1}{N}\sum_k
    \Bigl[
     {\bar  \rho}_{12}^z({\mit\Gamma};k){\rm cos}\frac{k}{3}
    -{\tilde\rho}_{12}^z({\mit\Gamma};k){\rm sin}\frac{k}{3}
    \Bigr]\,,
  \nonumber \\
  &&
  t_{1:n;3:n}^z=t_{2:n;3:n}^z
  \nonumber \\
  &&\quad
   =\frac{1}{N}
    \sum_k
    \Bigl[
     {\bar  \rho}_{13}^z({\mit\Gamma};k){\rm cos}\frac{k}{3}
    +{\tilde\rho}_{13}^z({\mit\Gamma};k){\rm sin}\frac{k}{3}
    \Bigr]\,.
  \nonumber
\end{eqnarray}
%\begin{eqnarray}
%  &&
%  s_{1:n}^z=s_{2:n}^z
%   =\frac{1}{N}
%    \sum_k
%    {\bar\rho}_{11}^z({\mit\Gamma};k)\,,\ \ 
%  s_{3:n}^z
%   =\frac{1}{N}
%    \sum_k
%    {\bar\rho}_{33}^z({\mit\Gamma};k)\,,
%  \nonumber \\
%  &&
%  t_{1:n;2:n-1}^z
%   =\frac{1}{N}\sum_k
%    \Bigl[
%     {\bar  \rho}_{12}^z({\mit\Gamma};k){\rm cos}\frac{k}{3}
%    -{\tilde\rho}_{12}^z({\mit\Gamma};k){\rm sin}\frac{k}{3}
%    \Bigr]\,,
%  \nonumber \\
%  &&
%  t_{1:n;3:n}^z=t_{2:n;3:n}^z
%   =\frac{1}{N}
%    \sum_k
%    \Bigl[
%     {\bar  \rho}_{13}^z({\mit\Gamma};k){\rm cos}\frac{k}{3}
%    +{\tilde\rho}_{13}^z({\mit\Gamma};k){\rm sin}\frac{k}{3}
%    \Bigr]\,.
%  \nonumber
%\end{eqnarray}

\noindent
(f) ${\mit\Gamma}B\otimes\check{S}^1\otimes\check{T}^1$
\begin{eqnarray}
  &&
  s_{1:n}^z=-s_{2:n}^z
   =\frac{1}{N}
    \sum_k
    {\bar\rho}_{11}^z({\mit\Gamma};k)\,,\ \ 
  s_{3:n}^z=0\,,
  \nonumber \\
  &&
  t_{1:n;2:n-1}^z=0\,,\ \ 
  t_{1:n;3:n}^z=-t_{2:n;3:n}^z
  \nonumber \\
  &&\quad
   =\frac{1}{N}
    \sum_k
    \Bigl[
     {\bar  \rho}_{13}^z({\mit\Gamma};k){\rm cos}\frac{k}{3}
    +{\tilde\rho}_{13}^z({\mit\Gamma};k){\rm sin}\frac{k}{3}
    \Bigr]\,.
  \nonumber
\end{eqnarray}
%\begin{eqnarray}
%  &&
%  s_{1:n}^z=-s_{2:n}^z
%   =\frac{1}{N}
%    \sum_k
%    {\bar\rho}_{11}^z({\mit\Gamma};k)\,,\ \ 
%  s_{3:n}^z=0\,,
%  \nonumber \\
%  &&
%  t_{1:n;2:n-1}^z=0\,,\ \ 
%  t_{1:n;3:n}^z=-t_{2:n;3:n}^z
%   =\frac{1}{N}
%    \sum_k
%    \Bigl[
%     {\bar  \rho}_{13}^z({\mit\Gamma};k){\rm cos}\frac{k}{3}
%    +{\tilde\rho}_{13}^z({\mit\Gamma};k){\rm sin}\frac{k}{3}
%    \Bigr]\,.
%  \nonumber
%\end{eqnarray}

\noindent
(g) $XA           \otimes\check{S}^1\otimes\check{T}^1$
\begin{eqnarray}
  &&
  s_{1:n}^z=s_{2:n}^z
  \nonumber \\
  &&\quad
   =\frac{(-1)^n}{N}
    \sum_k
    \Bigl[
     {\bar  \rho}_{11}^z(X;k){\rm cos}\frac{\pi}{3}
    -{\tilde\rho}_{11}^z(X;k){\rm sin}\frac{\pi}{3}
    \Bigr]\,,
  \nonumber \\
  &&
  s_{3:n}^z
   =\frac{(-1)^n}{N}
    \sum_k
    {\bar\rho}_{33}^z(X;k)\,,
  \nonumber \\
  &&
  t_{1:n;2:n-1}^z=0\,,\ \ 
  t_{1:n;3:n}^z=t_{2:n;3:n}^z
  \nonumber \\
  &&\quad
   =\frac{(-1)^n}{N}
    \sum_k
    \Bigl[
     {\bar  \rho}_{13}^z(X;k){\rm cos}\frac{k}{3}
    +{\tilde\rho}_{13}^z(X;k){\rm sin}\frac{k}{3}
    \Bigr]\,.
  \nonumber
\end{eqnarray}
%\begin{eqnarray}
%  &&
%  s_{1:n}^z=s_{2:n}^z
%   =\frac{(-1)^n}{N}
%    \sum_k
%    \Bigl[
%     {\bar  \rho}_{11}^z(X;k){\rm cos}\frac{\pi}{3}
%    -{\tilde\rho}_{11}^z(X;k){\rm sin}\frac{\pi}{3}
%    \Bigr]\,,\ \ 
%  s_{3:n}^z
%   =\frac{(-1)^n}{N}
%    \sum_k
%    {\bar\rho}_{33}^z(X;k)\,,
%  \nonumber \\
%  &&
%  t_{1:n;2:n-1}^z=0\,,\ \ 
%  t_{1:n;3:n}^z=t_{2:n;3:n}^z
%   =\frac{(-1)^n}{N}
%    \sum_k
%    \Bigl[
%     {\bar  \rho}_{13}^z(X;k){\rm cos}\frac{k}{3}
%    +{\tilde\rho}_{13}^z(X;k){\rm sin}\frac{k}{3}
%    \Bigr]\,.
%  \nonumber
%\end{eqnarray}

\noindent
(h) $XB           \otimes\check{S}^1\otimes\check{T}^1$
\begin{eqnarray}
  &&
  s_{3:n}^z=0\,,\ \ 
  s_{1:n}^z=-s_{2:n}^z
  \nonumber \\
  &&\quad
   =\frac{(-1)^n}{N}
    \sum_k
    \Bigl[
     {\bar  \rho}_{11}^z(X;k){\rm cos}\frac{\pi}{3}
    -{\tilde\rho}_{11}^z(X;k){\rm sin}\frac{\pi}{3}
    \Bigr]\,,
  \nonumber \\
  &&
  t_{1:n;2:n-1}^z=
   -\frac{(-1)^n}{N}
    \sum_k
  \nonumber \\
  &&\quad\times
    \Bigl[
     {\bar  \rho}_{12}^z(X;k){\rm cos}\frac{k-\pi}{3}
    -{\tilde\rho}_{12}^z(X;k){\rm sin}\frac{k-\pi}{3}
    \Bigr]\,,
  \nonumber \\
  &&
  t_{1:n;3:n}^z=-t_{2:n;3:n}^z
  \nonumber \\
  &&\quad
   =\frac{(-1)^n}{N}
    \sum_k
    \Bigl[
     {\bar  \rho}_{13}^z(X;k){\rm cos}\frac{k}{3}
    +{\tilde\rho}_{13}^z(X;k){\rm sin}\frac{k}{3}
    \Bigr]\,.
  \nonumber
\end{eqnarray}
%\begin{eqnarray}
%  &&
%  s_{3:n}^z=0\,,\ \ 
%  s_{1:n}^z=-s_{2:n}^z
%   =\frac{(-1)^n}{N}
%    \sum_k
%    \Bigl[
%     {\bar  \rho}_{11}^z(X;k){\rm cos}\frac{\pi}{3}
%    -{\tilde\rho}_{11}^z(X;k){\rm sin}\frac{\pi}{3}
%    \Bigr]\,,
%  \nonumber \\
%  &&
%  t_{1:n;2:n-1}^z=
%   -\frac{(-1)^n}{N}
%    \sum_k
%    \Bigl[
%     {\bar  \rho}_{12}^z(X;k){\rm cos}\frac{k-\pi}{3}
%    -{\tilde\rho}_{12}^z(X;k){\rm sin}\frac{k-\pi}{3}
%    \Bigr]\,,
%  \nonumber \\
%  &&
%  t_{1:n;3:n}^z=-t_{2:n;3:n}^z
%   =\frac{(-1)^n}{N}
%    \sum_k
%    \Bigl[
%     {\bar  \rho}_{13}^z(X;k){\rm cos}\frac{k}{3}
%    +{\tilde\rho}_{13}^z(X;k){\rm sin}\frac{k}{3}
%    \Bigr]\,.
%  \nonumber
%\end{eqnarray}
\end{appendix}

\widetext
\begin{table}
\caption{Symmetry operations of the group elements $g\in{\bf G}$ on
         the electron operators.}
\begin{tabular}{ccccc}
& $l$ & $C_2$ & $u({\bf e},\theta)$ & $t$ \\
\tableline
\noalign{\vskip 2pt}
$a_{1:k,s}^\dagger$ &
${\rm e}^{-{\rm i}k}a_{1:k,s}^\dagger$ &
$a_{2:-k,s}^\dagger$ &
$\sum_{s'}[u({\bf e},\theta)]_{ss'}
 a_{1:k,s'}^\dagger$ &
$(-1)^s a_{1:-k,-s}^\dagger$ \\
$a_{2:k,s}^\dagger$ &
${\rm e}^{-{\rm i}k}a_{2:k,s}^\dagger$ &
$a_{1:-k,s}^\dagger$ &
$\sum_{s'}[u({\bf e},\theta)]_{ss'}
 a_{2:k,s'}^\dagger$ &
$(-1)^s a_{2:-k,-s}^\dagger$ \\
$a_{3:k,s}^\dagger$ &
${\rm e}^{-{\rm i}k}a_{3:k,s}^\dagger$ &
$a_{3:-k,s}^\dagger$ &
$\sum_{s'}[u({\bf e},\theta)]_{ss'}
 a_{3:k,s'}^\dagger$ &
$(-1)^s a_{3:-k,-s}^\dagger$ \\
$a_{1:k+\pi,s}^\dagger$ &
$-{\rm e}^{-{\rm i}k}a_{1:k+\pi,s}^\dagger$ &
${\rm e}^{-(2\pi/3){\rm i}}a_{2:-k+\pi,s}^\dagger$ &
$\sum_{s'}[u({\bf e},\theta)]_{ss'}
 a_{1:k+\pi,s'}^\dagger$ &
$(-1)^s{\rm e}^{-(2\pi/3){\rm i}}a_{1:-k+\pi,-s}^\dagger$ \\
$a_{2:k+\pi,s}^\dagger$ &
$-{\rm e}^{-{\rm i}k}a_{2:k+\pi,s}^\dagger$ &
${\rm e}^{ (2\pi/3){\rm i}}a_{1:-k+\pi,s}^\dagger$ &
$\sum_{s'}[u({\bf e},\theta)]_{ss'}
 a_{2:k+\pi,s'}^\dagger$ &
$(-1)^s{\rm e}^{ (2\pi/3){\rm i}}a_{2:-k+\pi,-s}^\dagger$ \\
$a_{3:k+\pi,s}^\dagger$ &
$-{\rm e}^{-{\rm i}k}a_{3:k+\pi,s}^\dagger$ &
$a_{3:-k+\pi,s}^\dagger$ &
$\sum_{s'}[u({\bf e},\theta)]_{ss'}
 a_{3:k+\pi,s'}^\dagger$ &
$(-1)^s a_{3:-k+\pi,-s}^\dagger$ \\
\end{tabular}
\label{T:GA}
\end{table}
\end{document}